\documentclass[10pt, conference]{IEEEtran}

\usepackage[utf8]{inputenc}
\usepackage{amsmath, amssymb}
\usepackage{graphicx}
\usepackage{cite}
\usepackage{hyperref}
\usepackage{layout}
\usepackage[countmax]{subfloat}
\usepackage{float}
\usepackage[ruled]{algorithm2e}
\usepackage{multirow}
\usepackage{threeparttable}
\usepackage{booktabs}
\title{DYNAMO: Dynamic Neutral Atom Multi-programming Optimizer Towards Quantum Operating Systems}

\author{
	\IEEEauthorblockN{
		Wenjie~SUN\IEEEauthorrefmark{1},
		Xiaoyu~LI\IEEEauthorrefmark{2}\thanks{Corresponding author: xiaoyuuestc@uestc.edu.cn},
		Zhigang~Wang\IEEEauthorrefmark{1},
		Geng~CHEN\IEEEauthorrefmark{3},
		Lianhui~YU\IEEEauthorrefmark{3},
		Guowu~YANG\IEEEauthorrefmark{3}
	}
	\\
	\IEEEauthorblockA{\IEEEauthorrefmark{1}
		The School of Electronic Science and Engineering,\\ University of Electronic Science and Technology of China, Chengdu 611731, China}
	\\
	\IEEEauthorblockA{\IEEEauthorrefmark{2}The School of Information and Software Engineering,\\ University of Electronic Science and Technology of China, Chengdu 611731, China}
	\\
	\IEEEauthorblockA{\IEEEauthorrefmark{3}The School of Computer Science and Engineering,\\ University of Electronic Science and Technology of China, Chengdu 611731, China}
}

\begin{document}
	
\maketitle

\begin{abstract}
As quantum computing advances towards practical applications, quantum operating systems become inevitable, where multi-programming $\textendash$ the core functionality of operating systems $\textendash$ enables concurrent execution of multiple quantum programs to enhance hardware utilization. However, most quantum compilation work focuses solely on single-circuit execution, severely limiting resource efficiency and hindering quantum operating system development. We propose \textbf{D}ynamic \textbf{N}eutral \textbf{A}tom \textbf{M}ulti-programming \textbf{O}ptimizer (DYNAMO), a method that realizes multi-programming on neutral atom quantum architectures through parallel compilation and intelligent resource allocation across multiple quantum processing units (QPUs). DYNAMO addresses two critical challenges: inefficient and difficult resource partitioning, and complex scheduling conflicts from concurrent program. Our method enables efficient spatial and temporal resource sharing while maintaining circuit correctness and hardware constraints. Experimental evaluation across circuits ranging from $12$ to over $1200$ gates demonstrates that DYNAMO achieves up to $14.39\times$ compilation speedup while reducing execution stages by an average of $50.47\%$. Furthermore, DYNAMO successfully distributes workloads across multiple QPUs with balanced resource utilization. By enabling efficient multi-programming capabilities, DYNAMO establishes a critical foundation towards realizing practical quantum operating systems.
\end{abstract}

\section{Introduction}

Quantum computing promises exponential speedup for certain classes of problems, offering transformative potential across cryptography, optimization, and materials science \cite{Bernstein2017,Lancellotti2024,Abbas2024,Clinton2024,Harrigan2021,Arute2019,Huang2023}. As quantum computing transitions from theoretical exploration to practical implementation, the development of robust quantum processing units (QPUs) has become the cornerstone of realizing scalable quantum advantages. The maturation of quantum hardware platforms necessitates sophisticated software infrastructure to fully harness their computational capabilities \cite{Wu2013}.

Paralleling the evolution of classical computing systems, the emergence of quantum operating systems \cite{corrigan2017quantum} capable of scheduling and managing multiple QPUs \cite{acharya2024quantum} becomes inevitable. The core functionality enabling such systems is multi-programming \cite{Das2019,niu2023enabling}, which allows concurrent execution of multiple quantum programs to maximize hardware utilization and computational throughput. At a concrete level, the execution of a quantum program is the process of compiling qubits and running circuits, with the former being the key. Although existing compilation methods focus on optimizing individual quantum circuits, the growing demand for scalable quantum workloads highlights the critical importance of quantum multi-programming capabilities at the hardware level.

Series of work showing exploration of multi-programming capabilities. In 2019, Poulami Das et al. first proposed a quantum multi-programming method based on fair and reliable physical quantum bit resource partitioning, effectively improving the throughput and utilization of quantum computers\cite{Das2019}. In 2021, Liu et al. proposed QuCloud, which proposed a quantum multi-programming framework on superconducting quantum computers and proposed the use of X-SWAP scheme to improve program scheduling efficiency, laying foundations for quantum multi-programming\cite{Liu2021}. In 2022, Ohkura et al. took the crosstalk effect into consideration and proposed a crosstalk detection method, and improved the fidelity and quantum circuit execution time. \cite{Ohkura2022}. In 2024, Lei Liu et al. proposed QuCloud+, which took crosstalk into consideration in QuCloud's X-SWAP work, improved the fidelity, and reduced the depth of the compiled quantum circuit\cite{Liu2024}. In 2024, Orenstein et al. introduced QGroup, a parallel quantum job scheduling framework leveraging dynamic programming for superconducting quantum architectures, demonstrating improvements in execution parallelism and circuit fidelity. \cite{Orenstein2024}. In 2025, Li et al. proposed a multi-programming method based on genetic algorithm and corporate job splitting for superconducting quantum computers. For variational quantum eigensolver (VQE) circuits, this method achieved high fidelity and high throughput\cite{Li2025}.

However, all current work is only carried out on superconducting quantum computing systems. Among various hardware platforms, neutral atom quantum processors stand out due to their high parallelism and strong gate fidelities, making them compelling candidates for scalable quantum architectures. Compared with superconducting systems, research on multi-programming of neutral atom quantum computers can give full play to the parallel advantages  \cite{Evered2023,Sunami2025} of neutral atom quantum computing and make better use of quantum computing resources. In recent years, the development of neutral atom quantum computing has accelerated rapidly, driven by advances in trapping technologies, Rydberg-mediated interactions, and control protocols \cite{Henriet2020,Ebadi2021,Bluvstein2022,Evered2023,Bluvstein2024}. While neutral atom quantum computing is still in its early stages, the unique architectural features of these systems — such as dynamic qubit rearrangement and high parallelism — necessitate specialized quantum compilation techniques to effectively map abstract quantum algorithms onto hardware-level operations. Several recent works have explored compilation strategies tailored to neutral atom systems, including layered methods \cite{Baker2021}, heuristic search techniques such as Monte Carlo tree search \cite{Li2023}, and annealing-based qubit placement \cite{Patel2023}. Solver-based approaches have also been proposed, with Tan et al. introducing the first such method \cite{Tan2022}, later extended by DPQA to enhance both efficiency and scalability \cite{Tan2024}. PARALLAX further improved execution fidelity by refining annealing-based strategies \cite{Ludmir2024}.

Despite these advances in neutral atom compilation, a critical gap remains between the architectural advantages of neutral atom systems and the fundamental requirements for quantum operating systems. While neutral atom platforms possess the ideal characteristics for multi-programming — high parallelism, dynamic reconfigurability — existing compilation methods are inherently limited to single-circuit optimization, failing to bridge this gap. This limitation prevents the exploitation of neutral atom systems' full potential for concurrent program execution, which is essential for quantum operating system development. The disconnect between hardware capabilities and software infrastructure represents a fundamental bottleneck that must be addressed to realize practical quantum computing systems with efficient resource management and multi-programming capabilities.

To bridge this critical gap, we propose  \textbf{D}ynamic \textbf{N}eutral \textbf{A}tom \textbf{M}ulti-programming \textbf{O}ptimizer (DYNAMO), a method that realizes multi-programming capabilities on neutral atom quantum architectures. DYNAMO addresses the challenge of enabling concurrent quantum program execution by implementing parallel compilation and intelligent resource allocation across multiple quantum processing units. Our approach tackles the complex constraints inherent in neutral atom systems — particularly the global impact of AOD movement and scheduling conflicts — through novel spatial resource management and constraint-based scheduling techniques. By successfully implementing multi-programming on neutral atom platforms, DYNAMO establishes the foundation for quantum operating systems, further unlocking the potential of neutral atom architectures.

The main contributions of our work are as follows:
\begin{enumerate}
	
	\item {This work introduces the multi-programming method for neutral atom quantum architectures, addressing the core functionality required for quantum operating system development by enabling concurrent execution of multiple quantum programs on shared hardware resources.}
	\item {A two-component technical approach is developed, consisting of a spatial deformation model for dynamic resource management and a constraint-based scheduling method using SMT solvers. This approach effectively handles complex AOD movement constraints and scheduling conflicts inherent in multi-programming scenarios.}
	\item {Comprehensive experimental evaluation across circuits ranging from $12$ to over $1200$ gates demonstrates performance improvements. With the same compilation quality, DYNAMO achieves up to $14.39\times$ compilation speedup while reducing execution stages by an average of $50.47\%$, with balanced workload distribution across multiple quantum processing units.}
	
\end{enumerate}

\section{Background and Motivation}
\subsection{Preliminaries on Neutral Atom Compilation}
\label{Preliminaries}

Quantum circuits provide the standard framework for representing quantum algorithms, defining computations as sequences of gate operations on qubits. It translates high-level circuit descriptions into executable forms that respect the device’s physical limitations and operational constraints \cite{Baker2021, Khatri2019,Guo2024}. This process involves carefully assigning logical qubits to physical locations, scheduling gate operations to optimize performance and minimize errors, and coordinating their execution to adhere to architectural restrictions \cite{Liu2021,Lin2023}.

In neutral atom quantum computing, individual atoms serve as qubits and are manipulated using precisely controlled optical fields \cite{Baker2021}. The system employs two complementary types of traps: static traps created by spatial light modulators (SLM) that arrange qubits in fixed lattices, and mobile traps generated by acousto-optic deflectors (AOD) that enable dynamic qubit movement within the plane. Two-qubit gates are realized through global Rydberg laser excitation when pairs of qubits are positioned within the interaction range defined by the blockade radius rb, with computation proceeding in discrete steps known as Rydberg stages \cite{Tan2022,Tan2024a}.

Specifically, constrained by the hardware limitations of dynamically field-programmable neutral atom arrays, the compilation process must satisfy the following constraints\cite{Tan2022,Tan2024a}:
\begin{enumerate}
	\item{\textbf{Two-qubit gate execution constraint}: The two qubits executing a two-qubit gate must be sufficiently close to each other to establish quantum entanglement and realize the target two-qubit gate operation.}
	\item{\textbf{Two-qubit gate parallelization constraint}: When two qubits are executing a two-qubit gate operation, other qubits must maintain sufficient distance from these two qubits to ensure that additional two-qubit gate operations can be executed simultaneously without interference.}
	\item{\textbf{AOD movement constraint}: AOD traps can be repositioned through the movement of AOD tweezers, but multiple AOD tweezers must not create path crossings in the row or column directions during their movement.}
\end{enumerate}

These constraints create significant compilation challenges that distinguish neutral atom systems from other quantum architectures. Unlike superconducting systems where qubit interactions are typically local and connectivity is fixed, neutral atom systems require global consideration of AOD movement effects, making simple resource partitioning approaches ineffective. The dynamic nature of qubit positioning, combined with strict AOD ordering constraints, necessitates sophisticated compilation strategies that can coordinate complex spatial-temporal resource allocation while maintaining circuit correctness and hardware feasibility. For more detail, visualization of hardware mechanisms, constraint illustrations with specific examples, and step-by-step mapping procedures are provided in the ~\ref{Detailed Compilation}.

\subsection{Related Work in Quantum Multi-programming}

To solve the compilation difficulties of neutral atom qubit arrays, various methods have been proposed in recent years. In 2023, Li et al. developed a neutral atom compilation approach combining greedy algorithms with Monte Carlo tree search (MCTS) \cite{Li2023}. In the same year, Patel et al. introduced GRAPHINE, a framework that optimizes initial qubit placement through simulated annealing, though it does not support dynamic atom reconfiguration \cite{Patel2023}. Subsequently, Ludmir et al. (2025) proposed PARALLAX, which, compared to GRAPHINE, enables compilation based on dynamically field-programmable neutral atom arrays \cite{Ludmir2024}. In 2024, Lin et al. developed ZAC, a compiler specifically designed for compilation tasks in zoned neutral atom quantum architectures \cite{Lin2024}. More recently, Tan et al. (2025) introduced Enola, a compilation method based on simulated annealing and independent set algorithms that incorporates quantum gate fidelity into optimization objectives while maintaining excellent scalability \cite{Tan2025}. These methods primarily employ heuristic algorithms such as annealing and Monte Carlo search techniques, which offer rapid solution times but exhibit relatively limited optimization effectiveness.

An alternative class of neutral atom compilation methods relies on exact solver techniques. Solver-based approaches have been extensively applied to superconducting quantum computer compilation problems, demonstrating proven effectiveness \cite{Tan2020a, Tan2021, Molavi2022, Cong2023, Lin2023, Shaik2024}. In 2022, Tan et al. pioneered the discretization of neutral atom structure state spaces, transforming the compilation problem into a satisfiability modulo theories (SMT) problem by characterizing the intrinsic operating principles of reconfigurable neutral atom array hardware through constraint conditions, thereby extending solver-based methods to the neutral atom quantum compilation domain \cite{Tan2022}. Building upon this foundation, Tan et al. (2024) proposed an improved scheme called DPQA, which significantly reduces computational complexity through the integration of heuristic greedy algorithms \cite{Tan2024a}.

While these compilation methods have achieved significant improvements in optimizing individual quantum circuits, they share a fundamental limitation: all focus exclusively on single-circuit compilation scenarios. This constraint becomes increasingly problematic as quantum computing systems scale and require efficient resource utilization strategies. The limitations of current quantum compilation research and the demand for better QPU resource management motivate our work.

\subsection{Motivation for DYNAMO}
\label{Motivation}

\begin{figure*}[t]
	\centering
	\includegraphics[width=1.8\columnwidth, trim=14.5cm 9cm 14.5cm 9.2cm, clip]{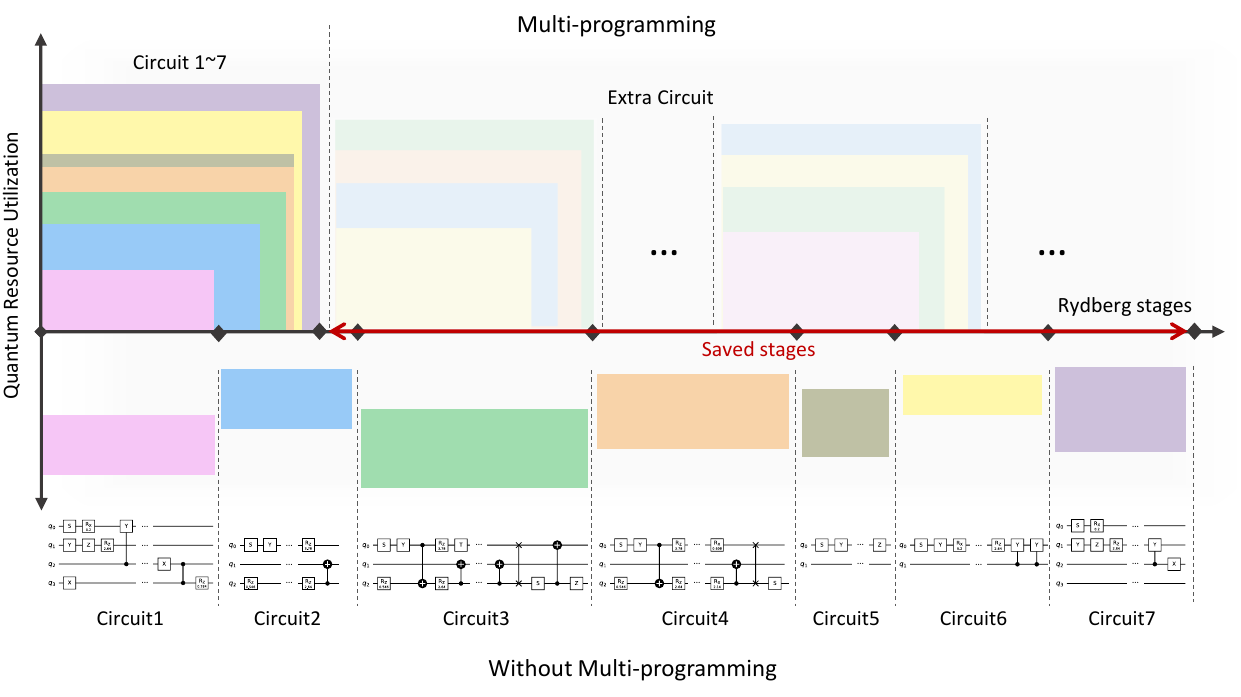}
	\caption{Comparison of Rydberg stages required and quantum resource utilization with and without multi-programming for circuit compilation. Different colored blocks represent the seven quantum circuits to be compiled. As shown in our preliminary experiments, multi-programming enables concurrent circuit execution, reducing compilation stages from $124$ to $32$ and improving quantum resource utilization.}
	\label{fig:utilization-and-stages}
\end{figure*}

Quantum multi-programming has gained significant attention as a fundamental approach to improve hardware utilization and enable quantum operating systems as shown in ~\ref{fig:utilization-and-stages}. Previous research has primarily focused on superconducting quantum computing platforms, demonstrating substantial improvements in throughput and resource efficiency through concurrent program execution. To the best of our knowledge, these efforts have not extended to neutral atom architectures, despite their superior characteristics for multi-programming applications.

The inherent parallelism of neutral atom systems — manifested through simultaneous AOD operations, concurrent gate executions within blockade constraints, and flexible spatial reconfiguration — creates a more desirable foundation for multi-programming implementations. Unlike fixed-connectivity quantum architectures, neutral atom systems can dynamically adapt their operational space to accommodate multiple concurrent programs, potentially achieving superior resource utilization compared to traditional approaches.

However, realizing multi-programming on neutral atom platforms requires addressing the complex interplay between the three fundamental compilation constraints and concurrent program execution demands. The global impact of AOD movement constraints, parallelization requirements within blockade radii, and spatial coordination challenges become significantly more complex when multiple programs compete for shared hardware resources, creating new optimization challenges that existing single-circuit methods cannot address.

\begin{figure}[h]
	\centering
	\includegraphics[width=1\columnwidth, trim=10.5cm 8cm 10.5cm 8.2cm, clip]{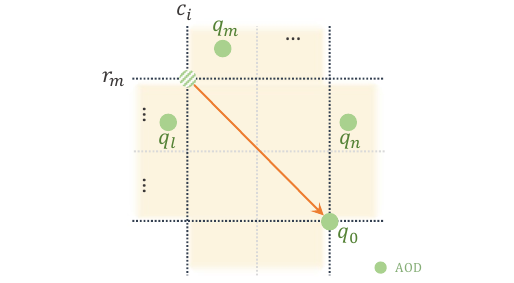}
	\caption{Global impact of moving $q_0$ in dynamically field-programmable neutral atom arrays, with affected regions and qubits highlighted in orange.}
	\label{fig:move-influence}
\end{figure}

To unlock the multi-programming potential of neutral atom architectures, two critical challenges must be addressed. \textbf{First}, the global nature of AOD movement constraints prevents simple resource partitioning strategies, necessitating sophisticated spatial resource management that can handle dynamic space deformation caused by concurrent program execution. \textbf{Second}, complex scheduling conflicts arise when multiple programs simultaneously compete for quantum resources, requiring intelligent constraint-based approaches that maintain circuit correctness while optimizing overall system performance. As demonstrated in ~\ref{fig:move-influence}, each AOD movement move affects a broad area, one small change triggers a chain reaction. Addressing these challenges is essential for realizing efficient multi-programming capabilities and establishing the foundation for quantum operating systems on neutral atom platforms.

\section{Parallel Compilation Framework for Dynamically Field - Programmable Neutral Atom Arrays}

As described in \ref{Motivation}, in order to address global AOD movement constraints and complex scheduling conflicts in multi-programming scenarios, a new approach to quantum compilation needs to be proposed. Traditional single-circuit optimization methods cannot address the spatial-temporal coordination required for concurrent program execution on neutral atom platforms. This motivates the development of a specialized parallel compilation method for dynamically field-programmable neutral atom arrays that can effectively manage multiple quantum programs while leveraging the inherent parallelism of these systems.

Our method draws inspiration from classical heterogeneous computing systems, where task scheduling across CPUs and GPUs has been optimized to improve utilization and throughput under constrained resources \cite{Pattnaik2016,Margiolas2016}, particularly job prioritization, cost-aware resource mapping, and bin-packing strategies\cite{Tillenius2015,Rey2016,CohenMaxime2017}. However, quantum compilation presents fundamentally different characteristics that require reversed prioritization strategies. In classical systems, longer jobs are often prioritized to reduce fragmentation since computational resources can be allocated independently. In contrast, quantum circuits must always be compiled from time t = 0 and potentially overlap with previously compiled circuits on the same array. Longer circuits occupy more future timesteps and create inflexible spatial-temporal obstacles that constrain subsequent circuit placement, leading to increased insertion delays and higher spatial contention.

Scheduling shorter circuits first leverages their limited temporal duration, allowing them to interfere only with the initial portions of later circuits while leaving substantial temporal freedom for longer programs. This asymmetric approach enables more efficient spatial utilization and reduces overall compilation complexity, as shorter circuits create smaller and more manageable spatial-temporal footprints that facilitate subsequent program placement. Based on these considerations, our method operates on two fundamental principles that guide the compilation process. \textbf{First}, shorter circuits should be scheduled earlier to minimize interference and enhance the schedulability of longer tasks. \textbf{Second}, workloads should be distributed across available quantum arrays to prevent local congestion and resource underutilization, ensuring balanced system performance.

We are given a set of $M$ quantum circuits $\mathcal{C} = \{C_1, C_2, \dots, C_M\}$ to be deployed on $N$ dynamically field-programmable neutral atom arrays $\mathcal{A} = \{A_1, A_2, \dots, A_N\}$. Each circuit $C_i$ is represented as a two-dimensional shape: its length corresponds to the number of Directed Acyclic Graph (DAG) layers, and its width at each step is the width (gate number) of that layer. Each quantum array $A_j$ supports a fixed spatial capacity, denoted as $W_\text{max}$, representing the maximum number of parallel qubits that can be operated at any given time. 

Our objective is to assign each circuit to an appropriate array and determine feasible insertion times such that spatial capacity constraints are respected at all timesteps while minimizing the total execution span across all arrays. This optimization balances compilation efficiency with resource utilization requirements. While global optimality would be ideal, the computational complexity of multi-circuit placement motivates a practical approach that prioritizes scalability and efficiency. We employ a fast, greedy heuristic that achieves  results with practical value while maintaining reasonable computational overhead for real-world deployment scenarios.To achieve these objectives, our scheduling method operates through a systematic two-phase process as follow:

(a) \textbf{The initial allocation phase} sorts all circuits in ascending order of length and directly assigns the shortest $N$ circuits to the $N$ arrays, one per array, starting from time $t = 0$. This phase establishes balanced baseline resource distribution and ensures that the most flexible (shortest) circuits occupy minimal temporal footprints, creating optimal conditions for subsequent placements.

(b) \textbf{The incremental assignment phase} then processes each remaining circuit $C_k$ in ascending order of length, simulating its insertion on each array $A_j$ by attempting placement starting from $t = 0$ and sliding the starting time forward until all timestep widths satisfy the spatial capacity constraint $W_{max}$. For each circuit, the placement that results in the smallest estimated post-placement array span is selected, effectively balancing array workloads while reduce overall system execution time.

(c) \textbf{Intra-array refinement phase} is applied after all circuits have been assigned, we perform a local refinement within each array $A_j$. Specifically, the circuits already placed on $A_j$ are re-sorted to further optimize execution order. The sorting follows a lexicographic two-tiered criterion: (1) circuits are first ordered by their earliest feasible start time (i.e., the earliest $t$ at which the circuit could have been legally inserted without violating the width constraint); (2) if multiple circuits share the same earliest start time, they are then ordered by ascending circuit length. This refinement enhances intra-array temporal efficiency and may expose additional opportunities for reducing idle time or enabling better synchronization across arrays.

This three-phase procedure ensures that short circuits gain early access to the schedule, that long circuits are placed with minimal disruption, and that the final schedule benefits from localized optimization within each array. The complete process is summarized in Algorithm\ref{alg:placement}.

\begin{algorithm}[t]
	\small
	\linespread{1.1}\selectfont
	\caption{Greedy Scheduling for Quantum Circuit Placement}
	\label{alg:placement}
	\KwIn{
		A set of quantum circuits $\mathcal{C} = \{C_1, \dots, C_M\}$,\\
		Number of arrays $N$, spatial capacity $W_\text{max}$
	}
	\KwOut{Mapping of each circuit to a quantum array and start time}
	\vspace{0.5em}
	
	\textbf{Step 1: Initialization}\\
	Sort all circuits in ascending order of length\;
	Assign the first $N$ circuits to arrays $A_1$ to $A_N$ at $t=0$\;
	Initialize each array’s timeline with the assigned circuit\;
	
	\vspace{0.5em}
	\textbf{Step 2: Incremental Placement}\\
	\ForEach{remaining circuit $C_k$} {
		\ForEach{array $A_j$} {
			$t \gets 0$\;
			\While{placement of $C_k$ at $t$ on $A_j$ violates width constraint} {
				$t \gets t + 1$\;
			}
			$\text{placement\_cost}(C_k, A_j) \gets t + \text{length}(C_k)$\;
		}
		Select $A_{j^*} = \arg\min_j \text{placement\_cost}(C_k, A_j)$\;
		Assign $C_k$ to $A_{j^*}$ starting at time $t$\;
		Update $A_{j^*}$'s timeline\;
	}
	
	\vspace{0.5em}
	\textbf{Step 3: Intra-Array Refinement Phase}\\
	\ForEach{array $A_j$} {
		Compute earliest feasible start time for each assigned circuit\;
		Sort circuits on $A_j$ by (i) ascending feasible start time, and (ii) ascending length in case of ties\;
		Update $A_j$'s timeline based on new order\;
	}
	
	\Return{All circuit-to-array mappings and start times}
\end{algorithm}

\section{Multi-programming Compilation Method}
\label{Multi-programming Compilation Method}

Based on the Parallel Compilation Framework, multiple quantum circuits are assigned to available quantum computing resources along with their respective execution orders to improve concurrency and throughput. However, for each individual neutral atom array resource (QPUs), we still require a specialized strategy to overcome the AOD movement constraint, and to more effectively harness the spatial concurrency potential of neutral atom platforms .

After a quantum circuit has been compiled for execution on a neutral atom array, the position and timing of each qubit’s operations are fully determined. Leveraging this, we introduce a cycle-wise decomposition of the compiled circuit, where each cycle is further subdivided into two ordered steps: (1) AOD Movement, and (2) Non-Movement Operations (e.g., gate execution, measurements). The entire compiled quantum program can then be represented as a sequential workflow, alternating between AOD Movement and Non-Movement Ops.

\begin{figure}[H]
	\centering
	\includegraphics[width=0.89\columnwidth, trim=10.5cm 7.5cm 8cm 6.4cm, clip]{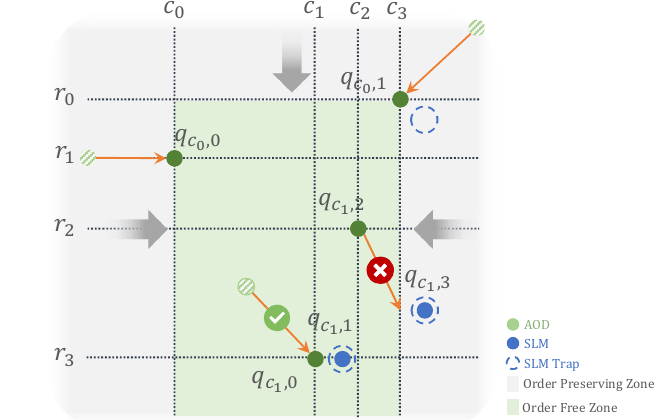}
	\caption{AOD movement in cycle 0.}
	\label{fig:multi-programming-0}
\end{figure}

\begin{figure}[H]
	\centering
	\includegraphics[width=0.89\columnwidth, trim=10.5cm 7.5cm 8cm 6.4cm, clip]{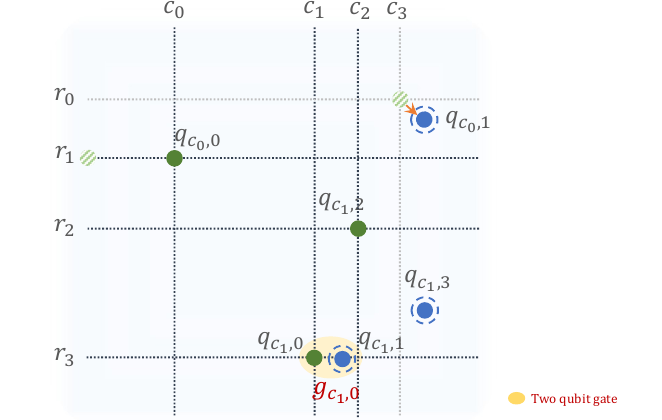}
	\caption{Atom transfer and two-qubit gate conduction in non-movement operations of cycle 0.}
	\label{fig:multi-programming-1}
\end{figure}

\begin{figure}[H]
	\centering
	\includegraphics[width=0.89\columnwidth, trim=10.5cm 7.5cm 8cm 6.4cm, clip]{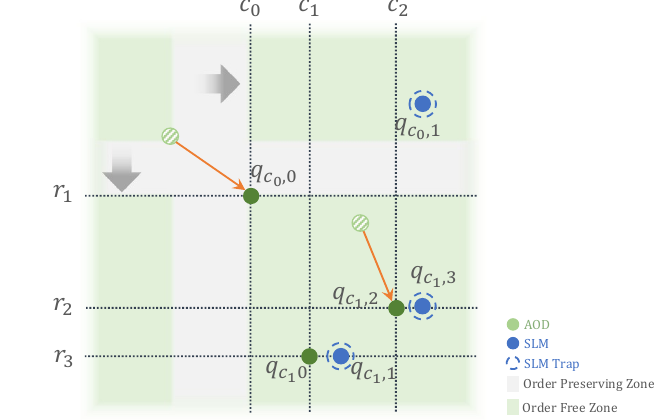}
	\caption{AOD movement in cycle 1.}
	\label{fig:multi-programming-2}
\end{figure}

\begin{figure}[H]
	\centering
	\includegraphics[width=0.89\columnwidth, trim=10.5cm 7.5cm 8cm 6.4cm, clip]{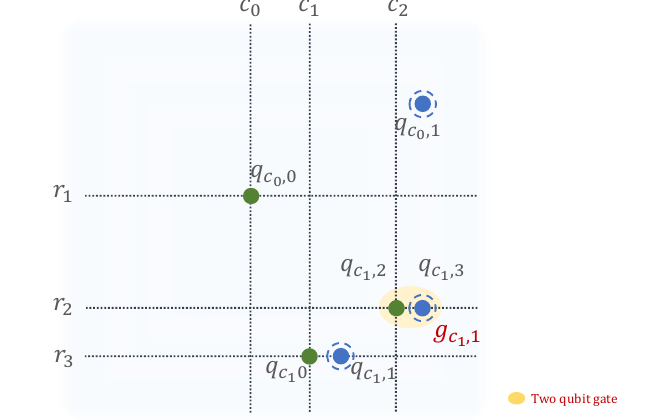}
	\caption{Atom transfer and two-qubit gate conduction in cycle 1.}
	\label{fig:multi-programming-3}
\end{figure}

This modeling approach explicitly separates global atom reconfigurations (AOD movements) from local gate-level operations, thus offering a clearer framework for implementing multi-programming on shared hardware. Importantly, our modeling does not alter the the original quantum compilation but instead serves as a structured interpretation of post-compilation behavior. To illustrate this approach, we present a concrete example demonstrating how spatial deformation enables multi-programming execution in in Figure \ref{fig:multi-programming-0} to Figure \ref{fig:multi-programming-3}. For ease of presentation, we denote two consecutive entire compiled quantum programs as cycle 0 and cycle 1, each complete AOD Movement and Non-Movement Ops, respectively.

In Figure \ref{fig:multi-programming-0}, a compiled quantum circuit $qc_0$ occupies two qubits $q_{c_0,0}$ and $q_{c_0,1}$, executing respective AOD movements during cycle 0. These movements, defined by their traversal of AOD rows/columns (e.g., row $r_0$, column $c_3$), partition the array into directional constraint regions which we term the Order-Preserving Zone (OPZ). These zones enforce movement constraints on any additional AOD row/column as follows:

\textbf{\textit{(a). AOD Row Constraint}}

Suppose an determined AOD row moves from position $y = y_0$ to $y = y_1$ within a given cycle. To ensure conflict-free execution, any \textbf{other} AOD row that is scheduled concurrently must satisfy the following spatial constraint:
\begin{itemize}
	\item If its initial position $y < y_0$, then its final position must satisfy $y < y_1$;
	\item If its initial position $y > y_0$, then its final position must satisfy $y > y_1$.
\end{itemize}

\textbf{\textit{(b). AOD Column Constraint}}

Similarly, suppose an AOD column moves from $x = x_0$ to $x = x_1$. Then, for any other concurrently moving AOD column:
\begin{itemize}
	\item If its initial position $x < x_0$, then its final position must satisfy $x < x_1$;
	\item If its initial position $x > x_0$, then its final position must satisfy $x > x_1$.
\end{itemize}

These constrained regions are annotated as grey in the figures, with directional vectors indicating allowable movement. The remaining regions unaffected by any movement constraints are termed Order-Free Zones (OFZ) and are marked green.

Based on the zonal layout of the neutral atom array, we now attempt to compile a second quantum circuit, denoted as $qc_1$. This circuit consists of four qubits, $q_{c_1,0}$, $q_{c_1,1}$, $q_{c_1,2}$, and $q_{c_1,3}$, and two two-qubit gates: $g_{c_1,0}$ acting on ($q_{c_1,0}$, $q_{c_1,1}$) and $g_{c_1,1}$ acting on ($q_{c_1,2}$, $q_{c_1,3}$). Due to the AOD constraints introduced by the previously compiled circuit $qc_0$ in cycle 0, the movement of $q_{c_1,2}$ toward $q_{c_1,3}$ is blocked because such movement would require AOD column $c_2$ to cross $c_3$, which violates the Order-Preserving constraint. In contrast, the movement of $q_{c_1,0}$ is not restricted since both its source and destination locations lie entirely within the Order-Free Zone and do not intersect any Order-Preserving Zone.

As a result, during the cycle 0 AOD movement step, $q_{c_1,0}$ moves adjacent to $q_{c_1,1}$, and the corresponding two-qubit gate $g_{c_1,0}$ is executed in the cycle 0 non-movement operation step, as illustrated in Figure~\ref{fig:multi-programming-1}. At this point, the gate $g_{c_1,1}$ remains unexecuted.

Figure~\ref{fig:multi-programming-2} depicts the cycle 1 AOD movement step, in which the movement of $qc_0$’s qubit $q_{c_0,0}$ occurs but does not interfere with the movement of $q_{c_1,2}$ toward $q_{c_1,3}$.

Finally, the two-qubit gate $g_{c_1,1}$ is executed in the cycle 1 non-movement operation step, as shown in Figure~\ref{fig:multi-programming-3}.

The key insight from this example lies in the dynamic partitioning of the compilation space. At each cycle, the presence of compiled circuits creates spatial deformation, yielding dynamic partitioning into OPZ and OFZ. AOD movements that traverse OPZs must comply with directional constraints established by existing circuit operations, whereas movements confined to OFZs remain unconstrained. This spatial deformation model enables multi-programming by allowing new circuits to be mapped onto the dynamically reshaped compilation space.

As guiding principles, the multi-programming compilation process requires satisfying AOD movement constraints that prevent spatial conflicts during concurrent execution, and two-qubit gate parallelization constraints that ensure spatial exclusivity of gate operations across all simultaneously executing circuits. To implement this spatial deformation approach systematically, we formalize the constraint relationships between concurrent programs in ~\ref{Multi-programming Compilation Details}. On top of this formulation, we design a constraint-based scheduler that incorporates array zoning and step placement logic to support concurrent compilation.

In summary, DYNAMO addresses multi-programming challenges through two complementary components. The parallel compilation framework employs a two-phase scheduling algorithm based on circuit length prioritization to distribute multiple quantum circuits across available quantum processing units. The multi-programming compilation method utilizes a spatial deformation model combined with constraint-based SMT scheduling to handle concurrent execution within individual processing units, managing AOD movement constraints and gate parallelization requirements through cycle-wise decomposition and dynamic partitioning. Together, these components enable the transition from single-circuit optimization to multi-programming capabilities on neutral atom platforms, contributing to the foundation for quantum operating systems. To evaluate DYNAMO's effectiveness, we conducted comprehensive experimental assessments across various circuit scales and compilation scenarios.

\section{Experimentation and Evaluation}

\subsection{Evaluation Framework and Baselines}

To evaluate the effectiveness and efficiency of our multi-programming compilation framework, we adopt two key metrics.

The first is the Rydberg stage count, which denotes the number of Rydberg stages required to execute the compiled quantum circuit. This metric is analogous to circuit depth in superconducting quantum architectures and serves as an indicator of quantum resource utilization and fidelity preservation \cite{Tan2020a,Tan2021,Lin2023,Shaik2024}. In dynamically field-programmable qubit arrays, a lower stage count reflects more efficient use of the shared quantum hardware, as it implies reduced contention for global operations and better parallelization across multiple programs. We define the Reduction Percentage Ratio (RPR\textsubscript{i}), where i = c or s, to quantify the relative reduction in stage count achieved by DYNAMO compared to DPQAc or DPQAs. Formally,

\begin{equation}
	\text{RPR}_{i} = \frac{\text{L}_\text{DYNAMO} - \text{L}_{\text{DPQA}^{i}}}{\text{L}_{\text{DPQA}^i}}
	\label{eq:RPR}
\end{equation}

The second metric is compilation time, defined as the total runtime of the compilation process \cite{Tan2022,Tan2024a}. This captures the practical efficiency of the method and reflects its scalability to large-scale or time-sensitive quantum workloads. Shorter compilation times are essential for responsive, on-demand compilation pipelines and for integration into hybrid quantum-classical workflows where rapid circuit deployment is critical. We define the $\text{Speedup}_{i}$ (where i = c or s) as the ratio of compilation time between DYNAMO and $\text{DPQA}^{\text{c/s}}$:

\begin{equation}
	\text{Speedup}_{i} = \frac{\text{T}_{\text{DYNAMO}}}{\text{T}_{\text{DPQA}^{i}}}
	\label{eq:Speedup}
\end{equation}

For baseline, we evaluate DYNAMO against DPQA\cite{Tan2024a}, the current state-of-the-art solver-based compilation method for neutral atom quantum systems. As DPQA has shown substantial performance gains over earlier techniques such as RAA\cite{Tan2022}, t$|\text{ket} \rangle$\cite{Sivarajah2020} and Sabre\cite{Li2019}, it serves as a strong and relevant baseline for comparison. Since DPQA natively supports only single-program compilation, we design two adaptation schemes to enable it to compile multiple quantum circuits for comparison purposes as sequential compilation scheme (denoted as $\text{DPQA}^s$) and merged compilation scheme (denoted as $\text{DPQA}^c$). The technical flow of the two compilation methods is shown in ~\ref{Experimental Details}

Benchmark circuits are sourced from RevLib \cite{Wille2008}, Sabre \cite{Li2019}, and QTetris \cite{Li2023}, covering a broad spectrum of circuit depths ranging from 12 to 3847. To mitigate the potential bias of larger circuits dominating the compilation process, we categorize the circuits into five distinct groups according to their depth as \textbf{Minimal} ($12 – 50$), \textbf{Minor} ($51 – 200$), \textbf{Moderate} ($201 – 800$), \textbf{Major} ($801 – 1,200$), and above \textbf{Maximal circuits} ($1,200+$).

\subsection{Experimental Design and Configuration}

Based on this benchmark set, we design three experiments to evaluate the performance of DYNAMO:

\begin{enumerate}
	\item{\textbf{Experiment 1}: Within each group, all circuits are paired to form two-circuit multi-program workloads. Each pair is compiled on a single neutral atom array using DYNAMO, $\text{DPQA}_s$, and $\text{DPQA}_c$. This experiment assesses DYNAMO’s compilation performance in small-scale multi-program scenarios.}
	\item{\textbf{Experiment 2}: All circuits within each group are compiled together as a grouped multi-program workload using DYNAMO, $\text{DPQA}_s$, and $\text{DPQA}_c$. This experiment measures DYNAMO’s effectiveness in compiling extended multi-program sequences on a single quantum array.}
	\item{\textbf{Experiment 3}: Each group is compiled using DYNAMO’s parallel framework by assigning the circuits to $2$ or $3$ separate neutral atom arrays. This evaluates the performance of DYNAMO in parallel compilation settings, highlighting its ability to scale with hardware resources.}
\end{enumerate}

A time limit of $10,000$ seconds is applied to each compilation task. These benchmarks and experiments offer a comprehensive and balanced evaluation across circuit sizes and compilation scenarios. More experimental details are shown in ~\ref{Experimental Details}.

\subsection{Pairwise Multi-Program Compilation}

\begin{table*}[h]
	\footnotesize
	\caption{Solving Time and Circuit Layers on Minimal and Minor Circuits}
	\label{tab:exp1-minimal-minor}
	\centering
	\tabcolsep=0.001\linewidth
	\renewcommand{\arraystretch}{1}
	\begin{tabular}{cccccccccccccc}
		\toprule
		\multirow{2}{*}{$\text{C}_{\text{map}}$} & 
		\multirow{2}{*}{$\text{C}_{\text{space}}$} &
		\multirow{2}{*}{$\text{Depth}_{\text{map}}$} & 
		\multirow{2}{*}{$\text{Depth}_{\text{space}}$} & 
		\multicolumn{5}{c}{\textbf{Solving time (s)}} & 
		\multicolumn{5}{c}{\textbf{Rydberg stages}} \\
		\cmidrule(lr){5-9}\cmidrule(lr){10-14} 
		& & & &
		$\text{T}_\text{DYNAMO}$ $^1$ & 
		$\text{T}_{\text{DPQA}^\text{c}}$ & 
		$\text{T}_{\text{DPQA}^\text{s}}$ & 
		$\text{Speedup}_{\text{c}}(\%)$ &  
		$\text{Speedup}_{\text{s}}(\%)$ &
		$\text{L}_\text{DYNAMO}$ & 
		$\text{L}_{\text{DPQA}^\text{c}}$ & 
		$\text{L}_{\text{DPQA}^\text{s}}$ & 
		$\text{RPR}_{\text{c}}(\%)$& 
		$\text{RPR}_{\text{s}}(\%)$ \\
		\midrule
		\multicolumn{14}{l}{\textbf{Minimal Circuits}} \\
		\addlinespace[0.5ex]
		4gt13\_92      & 4mod5-v1\_22   & 38 & 12 & 1.64  & 21.33 & 1.50  & 13.00 & 0.92 & 26 & 27 & 36 & $-3.70$  & $-27.78$ \\
		4gt13\_92      & alu-v0\_27     & 38 & 21 & 2.10  & 20.55 & 1.69  & 9.79  & 0.81 & 29 & 26 & 41 & 11.54    & $-29.27$ \\
		4gt13\_92      & bv\_n16        & 38 & 19 & 7.99  & 58.20 & 7.79  & 7.28  & 0.97 & 26 & 27 & 41 & $-3.70$  & $-36.59$ \\
		4gt13\_92      & decod24-v2\_43 & 38 & 30 & 1.85  & 18.59 & 1.70  & 10.05 & 0.92 & 26 & 26 & 48 & 0.00     & $-45.83$ \\
		4gt13\_92      & mod5mils\_65   & 38 & 21 & 2.09  & 21.39 & 1.75  & 10.24 & 0.84 & 27 & 26 & 42 & 3.85     & $-35.71$ \\
		4gt13\_92      & qv\_n16\_d5    & 38 & 36 & 10.82 & 65.23 & 10.04 & 6.03  & 0.93 & 35 & 29 & 46 & 20.69    & $-23.91$ \\
		qv\_n16\_d5    & 4mod5-v1\_22   & 36 & 12 & 11.16 & 54.18 & 9.38  & 4.85  & 0.84 & 24 & 25 & 30 & $-4.00$  & $-20.00$ \\
		qv\_n16\_d5    & alu-v0\_27     & 36 & 21 & 16.73 & 53.77 & 9.57  & 3.21  & 0.57 & 33 & 25 & 35 & 32.00    & $-5.71$  \\
		qv\_n16\_d5    & bv\_n16        & 36 & 19 & 18.38 & 55.51 & 15.66 & 3.02  & 0.85 & 24 & 26 & 35 & $-7.69$  & $-31.43$ \\
		qv\_n16\_d5    & decod24-v2\_43 & 36 & 30 & 9.57  & 54.09 & 9.58  & 5.65  & 1.00 & 22 & 25 & 42 & $-12.00$ & $-47.62$ \\
		qv\_n16\_d5    & mod5mils\_65   & 36 & 21 & 10.16 & 75.90 & 9.63  & 7.47  & 0.95 & 21 & 33 & 36 & $-36.36$ & $-41.67$ \\
		decod24-v2\_43 & 4mod5-v1\_22   & 30 & 12 & 1.10  & 15.83 & 1.04  & 14.39 & 0.95 & 22 & 22 & 32 & 0.00     & $-31.25$ \\
		decod24-v2\_43 & alu-v0\_27     & 30 & 21 & 1.52  & 15.87 & 1.23  & 10.42 & 0.81 & 28 & 22 & 37 & 27.27    & $-24.32$ \\
		decod24-v2\_43 & bv\_n16        & 30 & 19 & 7.43  & 15.67 & 7.33  & 2.11  & 0.99 & 22 & 22 & 37 & 0.00     & $-40.54$ \\
		decod24-v2\_43 & mod5mils\_65   & 30 & 21 & 1.60  & 17.71 & 1.29  & 11.10 & 0.81 & 29 & 22 & 38 & 31.82    & $-23.68$ \\
		alu-v0\_27     & 4mod5-v1\_22   & 21 & 12 & 1.13  & 12.17 & 1.03  & 10.76 & 0.91 & 15 & 15 & 25 & 0.00     & $-40.00$ \\
		alu-v0\_27     & bv\_n16        & 21 & 19 & 7.52  & 32.14 & 7.32  & 4.27  & 0.97 & 15 & 15 & 30 & 0.00     & $-50.00$ \\
		alu-v0\_27     & mod5mils\_65   & 21 & 21 & 1.86  & 12.53 & 1.28  & 6.75  & 0.69 & 23 & 16 & 31 & 43.75    & $-25.81$ \\
		mod5mils\_65   & 4mod5-v1\_22   & 21 & 12 & 1.19  & 12.85 & 1.09  & 10.84 & 0.92 & 16 & 16 & 26 & 0.00     & $-38.46$ \\
		mod5mils\_65   & bv\_n16        & 21 & 19 & 7.57  & 12.98 & 7.37  & 1.72  & 0.97 & 16 & 16 & 31 & 0.00     & $-48.39$ \\
		bv\_n16        & 4mod5-v1\_22   & 19 & 12 & 7.23  & 33.12 & 7.13  & 4.58  & 0.99 & 15 & 15 & 25 & 0.00     & $-40.00$ \\
		\midrule
		\multicolumn{7}{l}{\textbf{Average}} & \textbf{7.50}   & \textbf{0.89} & & & & \textbf{4.93}   & \textbf{-33.71} \\
		\midrule
		\addlinespace[1ex]
		\multicolumn{14}{l}{\textbf{Minor Circuits}} \\
		\addlinespace[0.5ex]
		rd84\_142        & ising\_model\_10 & 110 & 70  & 40.61 & 179.63 & 33.97 & 4.42 & 0.84 & 94  & 93  & 101 & 1.08     & $-6.93$  \\
		rd84\_142        & ising\_model\_13 & 110 & 71  & 42.65 & 195.01 & 36.32 & 4.57 & 0.85 & 92  & 98  & 101 & $-6.12$  & $-8.91$  \\
		rd84\_142        & ising\_model\_16 & 110 & 71  & 45.92 & 192.86 & 39.51 & 4.20 & 0.86 & 92  & 100 & 101 & $-8.00$  & $-8.91$  \\
		rd84\_142        & qft\_16          & 110 & 105 & 93.88 & 229.61 & 61.77 & 2.45 & 0.66 & 135 & 116 & 154 & 16.38    & $-12.34$ \\
		rd84\_142        & qv\_n12\_d10     & 110 & 71  & 50.81 & 127.11 & 39.83 & 2.50 & 0.78 & 101 & 92  & 119 & 9.78     & $-15.13$ \\
		qft\_16          & ising\_model\_10 & 105 & 70  & 36.36 & 164.83 & 34.33 & 4.53 & 0.94 & 71  & 75  & 93  & $-5.33$  & $-23.66$ \\
		qft\_16          & ising\_model\_13 & 105 & 71  & 40.77 & 200.46 & 36.68 & 4.92 & 0.90 & 74  & 91  & 93  & $-18.68$ & $-20.43$ \\
		qft\_16          & ising\_model\_16 & 105 & 71  & 42.44 & 202.43 & 39.87 & 4.77 & 0.94 & 71  & 92  & 93  & $-22.83$ & $-23.66$ \\
		qft\_16          & qv\_n12\_d10     & 105 & 71  & 53.46 & 157.31 & 40.19 & 2.94 & 0.75 & 90  & 98  & 111 & $-8.16$  & $-18.92$ \\
		qv\_n12\_d10     & ising\_model\_16 & 71  & 71  & 23.73 & 91.09  & 17.93 & 3.84 & 0.76 & 54  & 58  & 58  & $-6.90$  & $-6.90$  \\
		qv\_n12\_d10     & ising\_model\_10 & 71  & 70  & 13.76 & 58.43  & 12.39 & 4.25 & 0.90 & 40  & 38  & 58  & 5.26     & $-31.03$ \\
		qv\_n12\_d10     & ising\_model\_13 & 71  & 71  & 21.20 & 84.40  & 14.73 & 3.98 & 0.69 & 57  & 54  & 58  & 5.56     & $-1.72$  \\
		ising\_model\_16 & ising\_model\_13 & 71  & 71  & 28.69 & 94.02  & 14.41 & 3.28 & 0.50 & 47  & 45  & 40  & 4.44     & 17.50    \\
		ising\_model\_16 & ising\_model\_10 & 71  & 70  & 18.09 & 82.36  & 12.07 & 4.55 & 0.67 & 30  & 38  & 40  & $-21.05$ & $-25.00$ \\
		ising\_model\_13 & ising\_model\_10 & 71  & 70  & 13.38 & 66.34  & 8.87  & 4.96 & 0.66 & 32  & 39  & 40  & $-17.95$ & $-20.00$ \\
		\midrule
		\multicolumn{7}{l}{\textbf{Average}} & \textbf{4.01} & \textbf{0.78} & & & & \textbf{-4.83}  & \textbf{-13.74} \\
		\bottomrule
	\end{tabular}
	\begin{tablenotes}
		\footnotesize
		\item $^1$ Each QPU column contains the circuits assigned to that QPU, along with their compilation time $\Delta \mathrm{T}_{\mathrm{DYNAMO}}$ and the increase of post-compilation Rydberg stage counts $\Delta \mathrm{L}_{\mathrm{DYNAMO}}$.
	\end{tablenotes}
\end{table*}

Table \ref{tab:exp1-minimal-minor} presents the experimental results of Pairwise Multi-Program compilation on Minimal and Minor Circuits. In this and the following tables, $C_{space}$ refers to the initially mapped circuit, while $C_{map}$ denotes the subsequently mapped one. $Depth_{space}$ and $Depth_{map}$ represent the circuit depths, respectively. Compared with the merged DPQA approach, DYNAMO consistently achieves acceleration across all benchmarks, with a maximum speedup of $14.39\times$ and an average speedup of $7.5\times$. In terms of Rydberg stage count, DYNAMO performs comparably to merged DPQA, with an average increase of only $4.93\%$, indicating efficient scheduling. When compared with the sequential DPQA, DYNAMO shows slightly higher compilation time (about $1.12\times$ of sequential DPQA), yet significantly reduces the number of Rydberg stages in every case, achieving up to $50\%$ reduction, and $33.71\%$ reduction on average, highlighting its strong performance in space-time efficiency. Under the Minor Circuits Pairwise Multi-Program scenario, DYNAMO demonstrates consistent speedup over merged DPQA across all test cases, achieving a maximum speedup of $4.92\times$ and an average of $4.01\times$. Regarding Rydberg stage usage, DYNAMO maintains an advantage, with an average reduction of $4.83\%$, further reinforcing its spatial efficiency. While the compilation time of DYNAMO is slightly higher than that of sequential DPQA (about $1.28\times$), it nonetheless reduces the Rydberg stage count across all benchmarks, with up to $25\%$ and an average of $13.74\%$ stage reduction, showcasing its effectiveness in temporal resource savings.

\begin{table*}[h]
	\footnotesize
	\caption{Solving Time and Circuit Layers on Moderate Circuits}
	\label{tab:exp1-moderate}
	\centering
	\tabcolsep=0.0001\linewidth
	\renewcommand{\arraystretch}{1}
	\begin{tabular}{cccccccccccccc}
		\toprule
		\multirow{2}{*}{$\text{C}_{\text{map}}$} & 
		\multirow{2}{*}{$\text{C}_{\text{space}}$} &
		\multirow{2}{*}{$\text{Depth}_{\text{map}}$} & 
		\multirow{2}{*}{$\text{Depth}_{\text{space}}$} & 
		\multicolumn{5}{c}{\textbf{Solving time (s)}} & 
		\multicolumn{5}{c}{\textbf{Rydberg stages}} \\
		\cmidrule(lr){5-9}\cmidrule(lr){10-14} 
		& & & &
		$\text{T}_\text{DYNAMO}$ & 
		$\text{T}_{\text{DPQA}^\text{c}}$ & 
		$\text{T}_{\text{DPQA}^\text{s}}$ & 
		$\text{Speedup}_{\text{DPQA}^{\text{c}}}(\%)$ &  
		$\text{Speedup}_{\text{DPQA}^{\text{s}}}$(\%)&
		$\text{L}_\text{DYNAMO}$ & 
		$\text{L}_{\text{DPQA}^\text{c}}$ & 
		$\text{L}_{\text{DPQA}^\text{s}}$ & 
		$\text{RPR}_{\text{DPQA}^{\text{c}}}(\%)$&  
		$\text{RPR}_{\text{DPQA}^{\text{s}}}(\%)$ \\
		\midrule
		hwb5\_53    & cm152a\_212 & 758 & 684 & 184.26 & 493.17 & 155.64 & 2.68 & 0.84 & 777 & 542 & 998  & 43.36    & $-22.14$ \\
		hwb5\_53    & con1\_216   & 758 & 508 & 94.50  & 496.65 & 87.20  & 5.26 & 0.92 & 535 & 541 & 881  & $-1.11$  & $-39.27$ \\
		hwb5\_53    & f2\_232     & 758 & 668 & 98.11  & 503.72 & 89.60  & 5.13 & 0.91 & 535 & 536 & 985  & $-0.19$  & $-45.69$ \\
		hwb5\_53    & rd53\_130   & 758 & 569 & 79.61  & 537.85 & 72.74  & 6.76 & 0.91 & 535 & 544 & 918  & $-1.65$  & $-41.72$ \\
		hwb5\_53    & rd53\_251   & 758 & 712 & 105.59 & 607.59 & 98.15  & 5.75 & 0.93 & 535 & 542 & 1027 & $-1.29$  & $-47.91$ \\
		hwb5\_53    & sf\_274     & 758 & 436 & 73.36  & 492.35 & 58.45  & 6.71 & 0.80 & 673 & 538 & 835  & 25.09    & $-19.40$ \\
		hwb5\_53    & sf\_276     & 758 & 435 & 62.92  & 483.50 & 58.35  & 7.68 & 0.93 & 535 & 536 & 836  & $-0.19$  & $-36.00$ \\
		hwb5\_53    & wim\_266    & 758 & 514 & 119.12 & 780.64 & 110.91 & 6.55 & 0.93 & 535 & 539 & 887  & $-0.74$  & $-39.68$ \\
		rd53\_251   & cm152a\_212 & 712 & 684 & 223.05 & 567.01 & 175.16 & 2.54 & 0.79 & 748 & 505 & 955  & 48.12    & $-21.68$ \\
		rd53\_251   & con1\_216   & 712 & 508 & 114.73 & 546.63 & 106.72 & 4.76 & 0.93 & 492 & 494 & 838  & $-0.40$  & $-41.29$ \\
		rd53\_251   & f2\_232     & 712 & 668 & 118.28 & 546.35 & 109.12 & 4.62 & 0.92 & 492 & 502 & 942  & $-1.99$  & $-47.77$ \\
		rd53\_251   & rd53\_130   & 712 & 569 & 98.25  & 549.11 & 92.26  & 5.59 & 0.94 & 493 & 497 & 875  & $-0.80$  & $-43.66$ \\
		rd53\_251   & sf\_274     & 712 & 436 & 102.88 & 450.08 & 77.97  & 4.37 & 0.76 & 645 & 496 & 792  & 30.04    & $-18.56$ \\
		rd53\_251   & sf\_276     & 712 & 435 & 82.27  & 445.55 & 77.87  & 5.42 & 0.95 & 492 & 497 & 793  & $-1.01$  & $-37.96$ \\
		rd53\_251   & wim\_266    & 712 & 514 & 139.96 & 715.68 & 130.43 & 5.11 & 0.93 & 492 & 496 & 844  & $-0.81$  & $-41.71$ \\
		cm152a\_212 & con1\_216   & 684 & 508 & 173.04 & 565.07 & 164.22 & 3.27 & 0.95 & 461 & 484 & 809  & $-4.75$  & $-43.02$ \\
		cm152a\_212 & f2\_232     & 684 & 668 & 246.33 & 529.64 & 166.62 & 2.15 & 0.68 & 670 & 481 & 913  & 39.29    & $-26.62$ \\
		cm152a\_212 & rd53\_130   & 684 & 569 & 160.11 & 473.40 & 149.76 & 2.96 & 0.94 & 462 & 465 & 846  & $-0.65$  & $-45.39$ \\
		cm152a\_212 & sf\_274     & 684 & 436 & 193.50 & 418.29 & 135.47 & 2.16 & 0.70 & 625 & 471 & 763  & 32.70    & $-18.09$ \\
		cm152a\_212 & sf\_276     & 684 & 435 & 139.93 & 421.14 & 135.37 & 3.01 & 0.97 & 461 & 463 & 764  & $-0.43$  & $-39.66$ \\
		cm152a\_212 & wim\_266    & 684 & 514 & 199.73 & 652.78 & 187.93 & 3.27 & 0.94 & 461 & 463 & 815  & $-0.43$  & $-43.44$ \\
		f2\_232     & con1\_216   & 668 & 508 & 108.22 & 506.85 & 98.18  & 4.68 & 0.91 & 449 & 456 & 796  & $-1.54$  & $-43.59$ \\
		f2\_232     & rd53\_130   & 668 & 569 & 103.41 & 460.45 & 83.72  & 4.45 & 0.81 & 528 & 457 & 833  & 15.54    & $-36.61$ \\
		f2\_232     & sf\_274     & 668 & 436 & 86.56  & 411.76 & 69.43  & 4.76 & 0.80 & 528 & 456 & 750  & 15.79    & $-29.60$ \\
		f2\_232     & sf\_276     & 668 & 435 & 75.98  & 411.86 & 69.33  & 5.42 & 0.91 & 449 & 455 & 751  & $-1.32$  & $-40.21$ \\
		f2\_232     & wim\_266    & 668 & 514 & 133.27 & 642.22 & 121.89 & 4.82 & 0.91 & 449 & 456 & 802  & $-1.54$  & $-44.01$ \\
		rd53\_130   & con1\_216   & 569 & 508 & 89.77  & 381.19 & 81.32  & 4.25 & 0.91 & 384 & 387 & 729  & $-0.78$  & $-47.33$ \\
		rd53\_130   & sf\_274     & 569 & 436 & 75.15  & 348.93 & 52.57  & 4.64 & 0.70 & 542 & 388 & 683  & 39.69    & $-20.64$ \\
		rd53\_130   & sf\_276     & 569 & 435 & 57.47  & 346.02 & 52.47  & 6.02 & 0.91 & 383 & 387 & 684  & $-1.03$  & $-44.01$ \\
		rd53\_130   & wim\_266    & 569 & 514 & 114.13 & 558.75 & 105.03 & 4.90 & 0.92 & 383 & 387 & 735  & $-1.03$  & $-47.89$ \\
		wim\_266    & con1\_216   & 514 & 508 & 131.83 & 505.20 & 119.49 & 3.83 & 0.91 & 352 & 356 & 698  & $-1.12$  & $-49.57$ \\
		wim\_266    & sf\_274     & 514 & 436 & 142.11 & 499.10 & 90.74  & 3.51 & 0.64 & 515 & 354 & 652  & 45.48    & $-21.01$ \\
		wim\_266    & sf\_276     & 514 & 435 & 97.78  & 500.34 & 90.64  & 5.12 & 0.93 & 353 & 355 & 653  & $-0.56$  & $-45.94$ \\
		con1\_216   & sf\_274     & 508 & 436 & 102.40 & 306.90 & 67.03  & 3.00 & 0.65 & 512 & 348 & 646  & 47.13    & $-20.74$ \\
		con1\_216   & sf\_276     & 508 & 435 & 72.40  & 312.90 & 66.93  & 4.32 & 0.92 & 346 & 348 & 647  & $-0.57$  & $-46.52$ \\
		sf\_274     & sf\_276     & 436 & 435 & 42.37  & 269.73 & 38.18  & 6.37 & 0.90 & 301 & 305 & 601  & $-1.31$  & $-49.92$ \\
		\midrule
		\textbf{Average}     &             &     &     &        &        &        & \textbf{4.61} & \textbf{0.87} &     &     &      & \textbf{9.86}  & \textbf{-37.45} \\
		\bottomrule
	\end{tabular}
\end{table*}

Table \ref{tab:exp1-moderate} details the performance on Moderate Circuits using Pairwise Multi-Program compilation. Compared with merged DPQA, DYNAMO achieves acceleration across all circuits, reaching a maximum speedup of $6.76\times$ and an average of $4.61\times$. Although the Rydberg stage count is slightly higher (by an average of $9.86\%$), the trade-off is acceptable given the significant speed gains. Compared to the sequential DPQA, DYNAMO slightly increases the compilation time (about $1.15\times$), but again, consistently reduces Rydberg stages, with a maximum reduction of $49.92\%$ and an average of $37.45\%$.

\begin{table*}[h]
	\footnotesize
	\caption{Solving Time and Circuit Layers on Major and Maximal Circuits}
	\label{tab:exp1-major-maximal}
	\centering
	\tabcolsep=0.00001\linewidth
	\renewcommand{\arraystretch}{1}
	\begin{tabular}{cccccccccccccc}
		\toprule
		\multirow{2}{*}{$\text{C}_{\text{map}}$} & 
		\multirow{2}{*}{$\text{C}_{\text{space}}$} &
		\multirow{2}{*}{$\text{Depth}_{\text{map}}$} & 
		\multirow{2}{*}{$\text{Depth}_{\text{space}}$} & 
		\multicolumn{5}{c}{\textbf{Solving time (s)}} & 
		\multicolumn{5}{c}{\textbf{Rydberg stages}} \\
		\cmidrule(lr){5-9}\cmidrule(lr){10-14} 
		& & & &
		$\text{T}_\text{DYNAMO}$ & 
		$\text{T}_{\text{DPQA}^\text{c}}$ & 
		$\text{T}_{\text{DPQA}^\text{s}}$ & 
		$\text{Speedup}_{\text{DPQA}^{\text{c}}}(\%)$ &  
		$\text{Speedup}_{\text{DPQA}^{\text{s}}}(\%)$ &
		$\text{L}_\text{DYNAMO}$ & 
		$\text{L}_{\text{DPQA}^\text{c}}$ & 
		$\text{L}_{\text{DPQA}^\text{s}}$ & 
		$\text{RPR}_{\text{DPQA}^{\text{c}}}(\%)$&  
		$\text{RPR}_{\text{DPQA}^{\text{s}}}(\%)$ \\
		\midrule
		\multicolumn{14}{l}{\textbf{Major Circuits}} \\
		\addlinespace[0.5ex]
		squar5\_261 & cm42a\_207 & 1049 & 940  & 622.38 & 1299.25 & 444.41 & 2.09 & 0.71 & 1090 & 731 & 1354 & 49.11    & $-19.50$ \\
		squar5\_261 & dc1\_220   & 1049 & 1038 & 413.56 & 1306.69 & 377.90 & 3.16 & 0.91 & 720  & 737 & 1425 & $-2.31$  & $-49.47$ \\
		squar5\_261 & pm1\_249   & 1049 & 940  & 623.38 & 1312.36 & 442.32 & 2.11 & 0.71 & 1090 & 731 & 1354 & 49.11    & $-19.50$ \\
		dc1\_220    & cm42a\_207 & 1038 & 940  & 517.88 & 1069.21 & 382.52 & 2.06 & 0.74 & 1068 & 709 & 1339 & 50.63    & $-20.24$ \\
		dc1\_220    & pm1\_249   & 1038 & 940  & 514.15 & 1381.44 & 380.43 & 2.69 & 0.74 & 1068 & 730 & 1339 & 46.30    & $-20.24$ \\
		pm1\_249    & cm42a\_207 & 940  & 940  & 645.06 & 1220.55 & 446.94 & 1.89 & 0.69 & 989  & 655 & 1268 & 50.99    & $-22.00$ \\
		\midrule
		\multicolumn{7}{l}{\textbf{Average}} & \textbf{2.33} & \textbf{0.75} & & & & \textbf{40.64} & \textbf{-25.16} \\
		\midrule
		\addlinespace[1ex]
		\multicolumn{14}{l}{\textbf{Maximal Circuits}} \\
		\addlinespace[0.5ex]
		square\_root\_7 & adr4\_197       & 3847 & 1839 & 2894.90 & 6632.46 & 2333.54 & 2.29 & 0.81 & 3247 & 2554 & 3783 & 27.13    & $-14.17$ \\
		square\_root\_7 & cycle10\_2\_110 & 3847 & 3386 & 3895.49 & 7912.97 & 3037.90 & 2.03 & 0.78 & 3696 & 2608 & 4804 & 41.72    & $-23.06$ \\
		square\_root\_7 & misex1\_241     & 3847 & 2676 & 3325.35 & 7230.31 & 2857.05 & 2.17 & 0.86 & 3072 & 2557 & 4323 & 20.14    & $-28.94$ \\
		square\_root\_7 & radd\_250       & 3847 & 1781 & 2856.11 & 6446.18 & 2292.04 & 2.26 & 0.80 & 3242 & 2567 & 3737 & 26.30    & $-13.25$ \\
		square\_root\_7 & rd73\_252       & 3847 & 2867 & 3503.57 & 8573.91 & 2596.95 & 2.45 & 0.74 & 3668 & 2579 & 4489 & 42.23    & $-18.29$ \\
		square\_root\_7 & sym6\_145       & 3847 & 2187 & 2575.96 & 4233.18 & 2164.80 & 1.64 & 0.84 & 3072 & 2559 & 4025 & 20.05    & $-23.68$ \\
		square\_root\_7 & z4\_268         & 3847 & 1644 & 2669.50 & 5045.94 & 2168.85 & 1.89 & 0.81 & 3180 & 2545 & 3638 & 24.95    & $-12.59$ \\
		cycle10\_2\_110 & adr4\_197       & 3386 & 1839 & 1993.22 & 4734.71 & 1640.28 & 2.38 & 0.82 & 2978 & 2304 & 3537 & 29.25    & $-15.80$ \\
		cycle10\_2\_110 & misex1\_241     & 3386 & 2676 & 2321.96 & 6085.74 & 2163.79 & 2.62 & 0.93 & 2342 & 2314 & 4077 & 1.21     & $-42.56$ \\
		cycle10\_2\_110 & radd\_250       & 3386 & 1781 & 1933.41 & 5097.67 & 1598.78 & 2.64 & 0.83 & 2954 & 2343 & 3491 & 26.08    & $-15.38$ \\
		cycle10\_2\_110 & rd73\_252       & 3386 & 2867 & 2445.92 & 5064.85 & 1903.70 & 2.07 & 0.78 & 3359 & 2330 & 4243 & 44.16    & $-20.83$ \\
		cycle10\_2\_110 & sym6\_145       & 3386 & 2187 & 1835.28 & 3663.50 & 1471.55 & 2.00 & 0.80 & 3056 & 2303 & 3779 & 32.70    & $-19.13$ \\
		cycle10\_2\_110 & z4\_268         & 3386 & 1644 & 1764.35 & 4319.03 & 1475.59 & 2.45 & 0.84 & 2897 & 2304 & 3392 & 25.74    & $-14.59$ \\
		rd73\_252       & adr4\_197       & 2867 & 1839 & 1279.62 & 3481.12 & 1199.34 & 2.72 & 0.94 & 1964 & 2022 & 3222 & $-2.87$  & $-39.04$ \\
		rd73\_252       & misex1\_241     & 2867 & 2676 & 1860.77 & 3899.53 & 1722.85 & 2.10 & 0.93 & 1966 & 1998 & 3762 & $-1.60$  & $-47.74$ \\
		rd73\_252       & radd\_250       & 2867 & 1781 & 1429.25 & 3394.36 & 1157.84 & 2.37 & 0.81 & 2645 & 1991 & 3176 & 32.85    & $-16.72$ \\
		rd73\_252       & sym6\_145       & 2867 & 2187 & 1088.89 & 3080.89 & 1030.60 & 2.83 & 0.95 & 1965 & 2004 & 3464 & $-1.95$  & $-43.27$ \\
		rd73\_252       & z4\_268         & 2867 & 1644 & 1251.82 & 3630.58 & 1034.65 & 2.90 & 0.83 & 2577 & 2003 & 3077 & 28.66    & $-16.25$ \\
		misex1\_241     & adr4\_197       & 2676 & 1839 & 1553.87 & 4303.59 & 1459.43 & 2.77 & 0.94 & 1800 & 1828 & 3056 & $-1.53$  & $-41.10$ \\
		misex1\_241     & radd\_250       & 2676 & 1781 & 1892.24 & 3712.16 & 1417.93 & 1.96 & 0.75 & 2476 & 1834 & 3010 & 35.01    & $-17.74$ \\
		misex1\_241     & sym6\_145       & 2676 & 2187 & 1358.97 & 2692.47 & 1290.70 & 1.98 & 0.95 & 1800 & 1814 & 3298 & $-0.77$  & $-45.42$ \\
		misex1\_241     & z4\_268         & 2676 & 1644 & 1710.55 & 3288.59 & 1294.74 & 1.92 & 0.76 & 2429 & 1832 & 2911 & 32.59    & $-16.56$ \\
		sym6\_145       & adr4\_197       & 2187 & 1839 & 817.71  & 1979.72 & 767.18  & 2.42 & 0.94 & 1502 & 1518 & 2758 & $-1.05$  & $-45.54$ \\
		sym6\_145       & radd\_250       & 2187 & 1781 & 858.37  & 1986.27 & 725.68  & 2.31 & 0.85 & 2172 & 1509 & 2712 & 43.94    & $-19.91$ \\
		sym6\_145       & z4\_268         & 2187 & 1644 & 696.79  & 2555.89 & 602.49  & 3.67 & 0.86 & 2009 & 1517 & 2613 & 32.43    & $-23.12$ \\
		adr4\_197       & radd\_250       & 1839 & 1781 & 1234.96 & 2450.63 & 894.42  & 1.98 & 0.72 & 1929 & 1275 & 2470 & 51.29    & $-21.90$ \\
		adr4\_197       & z4\_268         & 1839 & 1644 & 1060.76 & 2205.83 & 771.23  & 2.08 & 0.73 & 1863 & 1299 & 2371 & 43.42    & $-21.43$ \\
		radd\_250       & z4\_268         & 1781 & 1644 & 1036.65 & 2060.51 & 729.73  & 1.99 & 0.70 & 1832 & 1236 & 2325 & 48.22    & $-21.20$ \\
		\midrule
		\multicolumn{7}{l}{\textbf{Average}} & \textbf{2.32} & \textbf{0.83} & & & & \textbf{25.01} & \textbf{-24.97} \\
		\bottomrule
	\end{tabular}
\end{table*}

Table \ref{tab:exp1-major-maximal} presents the results for Major Circuits. DYNAMO continues to outperform merged DPQA in all benchmarks, achieving a maximum speedup of $3.16\times$ and an average of $2.33\times$. While the average Rydberg stage count increases by $40.64\%$, this is accompanied by substantial improvements in compilation speed. When compared to the sequential DPQA, DYNAMO's runtime remains acceptable (about $1.33\times$ of sequential), while it delivers notable reductions in Rydberg stages for all circuits up to $49.47\%$ reduction and $25.16\%$ on average, demonstrating its effectiveness in reducing temporal depth despite increased spatial overhead. For the evaluation results on Maximal Circuits, DYNAMO surpasses merged DPQA in compilation speed, reaching a maximum speedup of $2.83\times$ and an average of $2.32\times$. Although the Rydberg stage count increases by $25.01\%$ on average, this is a reasonable trade-off given the parallel performance gains. When compared with sequential DPQA, DYNAMO incurs a moderate runtime increase (about $1.2\times$), yet consistently reduces the Rydberg stages, with a maximum reduction of $47.74\%$ and an average of $24.97\%$, confirming its strength in temporal compression of multi-program workloads.

Across all circuit groups ranging from Minimal to Maximal, the proposed DYNAMO framework consistently demonstrates substantial advantages in compilation efficiency and spatiotemporal performance. When compared to the merged DPQA baseline, DYNAMO achieves acceleration on all benchmarks, with speedup ratios reaching up to $14.39\times$, and averaging $4.36\times$ across circuit sizes. Despite a moderate increase in Rydberg stage counts in some cases — particularly for more complex circuits — DYNAMO maintains comparable or improved stage efficiency in most settings. Compared to the sequential DPQA approach, DYNAMO exhibits slightly longer compilation times (typically within $1.1–1.3\times$ range), yet consistently reduces the number of Rydberg stages by a significant margin, achieving average reductions of $50.47\%$. 

The pairwise multi-program compilation results demonstrate DYNAMO's effectiveness in small-scale multi-programming scenarios across diverse circuit complexities. It's worth noting that the performance improvements become more pronounced with larger circuits, suggesting that DYNAMO's spatial deformation model and constraint-based scheduling effectively handle the increased complexity of concurrent program execution. These results validate the fundamental approach of treating compiled circuits as dynamic space occupations, enabling effective resource sharing without compromising compilation correctness.

\subsection{Grouped Multi-Program Compilation}

To evaluate DYNAMO’s scalability in handling larger multi-program workloads, we compile all circuits within each group as unified task sets. Under a $10,000$-second time constraint, the merged DPQA baseline fails to complete compilation for any group, making sequential DPQA the primary baseline for comparison.

\begin{table*}[h]
	\footnotesize
	\caption{Multi-programming Compilation Results on All Circuit Categories}
	\label{tab:exp2-all}
	\centering
	\tabcolsep=0.2\linewidth
	\renewcommand{\arraystretch}{1.1}
	\begin{tabular}{@{}l@{\hspace{1em}}l@{\hspace{2em}}c@{\hspace{1.5em}}c@{\hspace{1em}}c@{\hspace{1.5em}}r@{\hspace{1.5em}}c@{\hspace{1em}}c@{\hspace{1em}}r@{}}
		\toprule
		& \textbf{Circuit} & \textbf{Depth} & \textbf{$\Delta \text{L}_\text{DYNAMO}$}$^1$ & \textbf{$\Delta \text{L}_{\text{DPQA}^{\text{s}}}$}  & \textbf{$\text{RPR}_{\text{s}}(\%)$} 
		& \textbf{$\Delta \text{T}_\text{DYNAMO}$}$^2$ & \textbf{$\Delta \text{T}_{\text{DPQA}^{\text{s}}}$} & \textbf{$\text{Speedup}_{\text{s}}(\%)$}\\
		\midrule
		\multirow{8}{*}{\rotatebox{90}{\textbf{Minimal}}} 
		& 4mod5-v1\_22   & 12           & 10          & 10           & 0.00            & 0.42           & 0.42           & 1             \\
		& bv\_n16        & 19           & 5           & 15           & -66.67          & 7.02           & 6.71           & 0.96          \\
		& alu-v0\_27     & 21           & 7           & 15           & -53.33          & 1.36           & 0.61           & 0.45          \\
		& mod5mils\_65   & 21           & 5           & 16           & -68.75          & 1.8            & 0.67           & 0.37          \\
		& decod24-v2\_43 & 30           & 8           & 22           & -63.64          & 1.58           & 0.62           & 0.39          \\
		& qv\_n16\_d5    & 36           & 24          & 20           & 20.00           & 34.59          & 8.96           & 0.26          \\
		& 4gt13\_92      & 38           & 0           & 26           & -100.00        & 4.5            & 1.08           & 0.24          \\
		\midrule
		& \textbf{Sum}   & \textbf{177} & \textbf{59} & \textbf{124} & \textbf{-52.42} & \textbf{51.27} & \textbf{19.06} & \textbf{0.37} \\
		\midrule
		\multirow{7}{*}{\rotatebox{90}{\textbf{Minor}}} 
		& ising\_model\_10 & 71           & 20           & 20           & 0.00            & 3.26            & 3.26           & 1             \\
		& ising\_model\_13 & 71           & 8            & 20           & -60.00          & 9.01            & 5.61           & 0.62          \\
		& ising\_model\_16 & 71           & 35           & 20           & 75.00           & 34.14           & 8.8            & 0.26          \\
		& qv\_n12\_d10     & 71           & 39           & 38           & 2.63            & 32.84           & 9.12           & 0.28          \\
		& qft\_16          & 105          & 44           & 73           & -39.73         & 86.17           & 31.07          & 0.36          \\
		& rd84\_142        & 110          & 47           & 81           & -41.98         & 104.8           & 30.71          & 0.29          \\
		\midrule
		& \textbf{Sum}     & \textbf{499} & \textbf{193} & \textbf{252} & \textbf{-23.41} & \textbf{270.22} & \textbf{88.57} & \textbf{0.33} \\
		\midrule
		\multirow{10}{*}{\rotatebox{90}{\textbf{Moderate}}} 
		& sf\_276      & 435           & 301           & 301           & 0.00            & 19.04            & 19.04           & 1             \\
		& con1\_216    & 508           & 53            & 346           & -84.68          & 54.83            & 47.89           & 0.87          \\
		& sf\_274      & 436           & 138           & 300           & -54.00          & 41.49            & 19.14           & 0.46          \\
		& wim\_266     & 514           & 150           & 352           & -57.39          & 173.05           & 71.6            & 0.41          \\
		& rd53\_130    & 569           & 104           & 383           & -72.85          & 95.52            & 33.43           & 0.35          \\
		& f2\_232      & 668           & 126           & 450           & -72.00          & 146.17           & 50.29           & 0.34          \\
		& cm152a\_212  & 684           & 84            & 463           & -81.86          & 353.53           & 116.33          & 0.33          \\
		& rd53\_251    & 712           & 78            & 492           & -84.15         & 197.25           & 58.83           & 0.3           \\
		& hwb5\_53     & 758           & 75            & 535           & -85.98          & 140.75           & 39.31           & 0.28          \\
		\midrule
		& \textbf{Sum} & \textbf{5284} & \textbf{1109} & \textbf{3622} & \textbf{-69.38} & \textbf{1221.63} & \textbf{455.86} & \textbf{0.37} \\
		\midrule
		\multirow{5}{*}{\rotatebox{90}{\textbf{Major}}} 
		& pm1\_249     & 940           & 634           & 634           & 0.00            & 222.42           & 222.42          & 1             \\
		& cm42a\_207   & 940           & 367           & 634           & -42.11          & 440.56           & 224.52          & 0.51          \\
		& dc1\_220     & 1038          & 284           & 705           & -59.72          & 384.98           & 158.01          & 0.41          \\
		& squar5\_261  & 1049          & 190           & 720           & -73.61          & 639.4            & 219.9           & 0.34          \\
		\midrule
		& \textbf{Sum} & \textbf{3967} & \textbf{1475} & \textbf{2693} & \textbf{-45.23} & \textbf{1687.36} & \textbf{824.84} & \textbf{0.49} \\
		\midrule
		\multirow{9}{*}{\rotatebox{90}{\textbf{Maximal}}} 
		& z4\_268         & 1644           & 1113          & 1113           & 0.00            & 303.27            & 303.27           & 1             \\
		& radd\_250       & 1781           & 726           & 1212           & -40.10         & 755.43            & 426.46           & 0.56          \\
		& adr4\_197       & 1839           & 476           & 1258           & -62.16          & 1026.62           & 467.96           & 0.46          \\
		& sym6\_145       & 2187           & 481           & 1500           & -67.93          & 562.47            & 299.22           & 0.53          \\
		& misex1\_241     & 2676           & 596           & 1798           & -66.85         & 2237.91           & 991.47           & 0.44          \\
		& rd73\_252       & 2867           & 532           & 1964           & -72.91          & 1550.15           & 731.38           & 0.47          \\
		& cycle10\_2\_110 & 3386           & 593           & 2279           & -73.98          & 2468.68           & 1172.32          & 0.47          \\
		& square\_root\_7 & 3847           & 683           & 2525           & -72.95          & 4264.95           & 1865.58          & 0.44          \\
		\midrule
		& \textbf{Sum}    & \textbf{20227} & \textbf{5200} & \textbf{13649} & \textbf{-61.90\%} & \textbf{13169.49} & \textbf{6257.66} & \textbf{0.48} \\
		\bottomrule
	\end{tabular}
	\begin{tablenotes}
		\footnotesize
		\item $^1$ $\Delta \mathrm{L}_{\mathrm{DYNAMO}}$ and $\Delta \mathrm{L}_{\mathrm{DPQA}^{\mathrm{s}}}$ denote the increase in Rydberg stage count after compiling the circuit on the left.
		\item  $^2$ $\Delta \mathrm{T}_{\mathrm{DYNAMO}}$ and $\Delta \mathrm{T}_{\mathrm{DPQA}^{\mathrm{s}}}$ indicate the increase in compilation time (in seconds) for the same circuit.
	\end{tablenotes}
\end{table*}

As shown in Table~\ref{tab:exp2-all}, across the Minimal, Minor, Moderate, Major, and Maximal circuit groups, DYNAMO requires $59$, $193$, $1109$, $1475$, and $5200$ Rydberg stages, respectively, to complete the same computational tasks. Compared to sequential DPQA, this represents quantum resource reductions of $52.4\%$, $23.8\%$, $68.4\%$, $45.2\%$, and $61.9\%$, respectively. These reductions become increasingly significant as the depth and complexity of the circuit groups grow, highlighting DYNAMO’s capability to manage spatiotemporal congestion and inter-program interaction overhead effectively. While DYNAMO generally incurs longer compilation times than sequential DPQA, the overhead remains within acceptable bounds for practical use.

\begin{figure*}[h]
	\centering
	\includegraphics[width=1\textwidth]{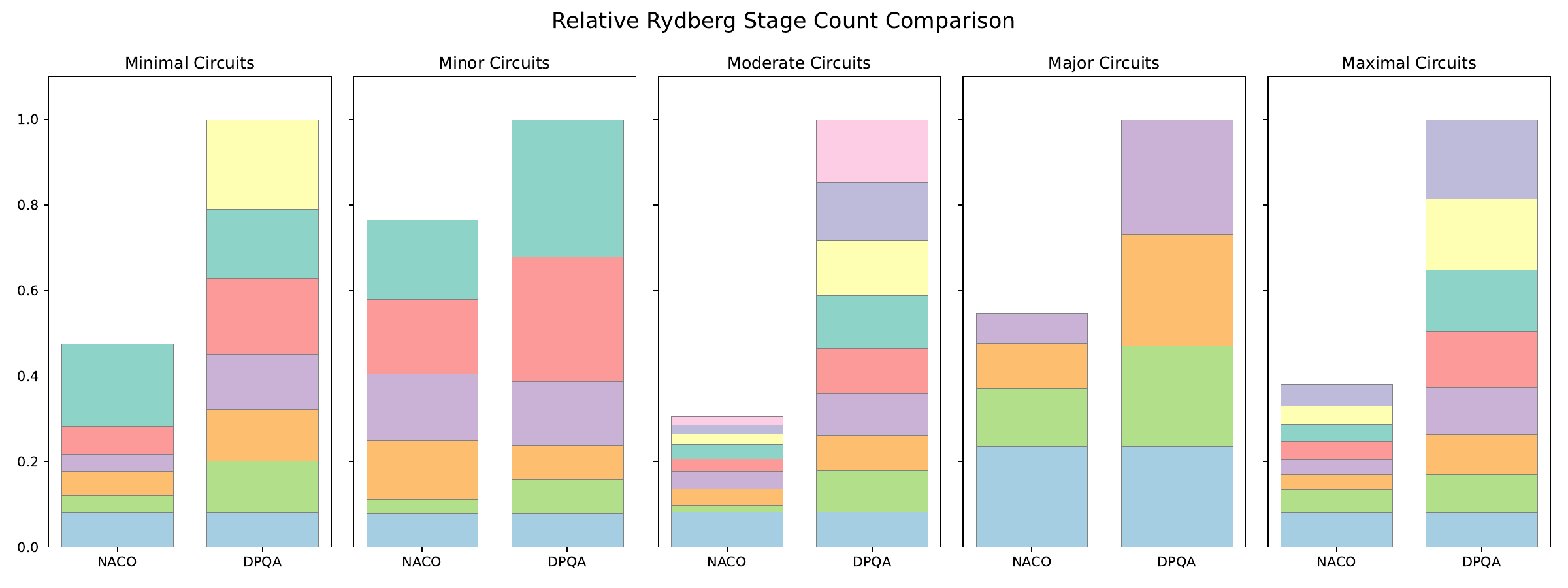}
	\caption{Normalized comparison of Rydberg stages between DYNAMO and sequential DPQA on five circuit groups. Each bar represents the total number of Rydberg stages normalized by the corresponding value obtained from sequential DPQA compilation. Within each bar, stacked color segments indicate the contributions of individual circuits in the group, reflecting their respective shares of Rydberg stages after compilation.}
	\label{fig:exp2total}
\end{figure*}

To further illustrate how such substantial reductions in execution stages are achieved, Figure~\ref{fig:exp2total} presents a normalized comparison of Rydberg stages per circuit between DYNAMO and sequential DPQA across all circuit groups. Notably, as the number of circuits within a group increases, the performance advantage of DYNAMO becomes more evident. For example, in the Moderate group (comprising 9 circuits) and the Maximal group (comprising 8 circuits), which include the largest number of programs, DYNAMO demonstrates significantly greater efficiency compared to smaller groups. This observation highlights DYNAMO’s effectiveness in handling large-scale multi-program compilation scenarios.

The efficiency gains of DYNAMO stem from its adaptive scheduling mechanism. Once a circuit is placed and executed, its spatiotemporal footprint implicitly expands the available window for subsequent programs. Unlike traditional methods that treat each circuit as an independent compilation layer, DYNAMO decomposes and strategically adapts circuit operations to occupy residual gaps left by prior executions — analogous to fitting irregular components into the remaining spaces of a mosaic. This dynamic reuse of both spatial and temporal resources enables compact and efficient integration of multiple quantum programs on reconfigurable neutral atom array. Notably, the performance advantages become more pronounced with larger circuit groups, as evidenced by the significant efficiency gains in Moderate and Maximal groups containing $9$ and $8$ circuits respectively. This scaling behavior validates our hypothesis that neutral atom architectures possess inherent parallelism advantages that can be effectively exploited through multi-programming approaches. The results strongly support the motivation for developing quantum operating systems capabilities, as DYNAMO successfully transforms the compilation process from resource-constrained single-circuit optimization to efficient concurrent program execution.

\subsection{Multi-Resource Multi-Program Compilation}

\begin{table*}[h]
	\footnotesize
	\caption{Multi-Program Compilation Results on Two Quantum Resources}
	\label{tab:exp2-qpu-all}
	\centering
	\tabcolsep=0.18\linewidth
	\renewcommand{\arraystretch}{1}
	\begin{tabular}{@{}l@{\hspace{1.5em}}l@{\hspace{1em}}r@{\hspace{1em}}r@{\hspace{2em}}l@{\hspace{1em}}r@{\hspace{1em}}r@{}}
		\toprule
		& 
		\multicolumn{3}{c}{\textbf{QPU 0}} & 
		\multicolumn{3}{c}{\textbf{QPU 1}} \\
		\cmidrule(lr){2-4}\cmidrule(lr){5-7}
		& \textbf{Circuit} & \textbf{$\Delta \text{L}_\text{DYNAMO}$} & \textbf{$\Delta \text{T}_\text{DYNAMO}$ $^1$} &
		\textbf{Circuit} & \textbf{$\Delta \text{L}_\text{DYNAMO}$} & \textbf{$\Delta \text{T}_\text{DYNAMO}$} \\
		\midrule
		\multirow{5}{*}{\rotatebox{90}{\textbf{Minimal}}} 
		& 4mod5-v1\_22   & 10 & 0.42 & bv\_n16      & 15 & 6.71  \\
		& decod24-v2\_43 & 12 & 0.67 & alu-v0\_27   & 8  & 1.38  \\
		& 4gt13\_92      & 12 & 1.75 & mod5mils\_65 & 4  & 1.68  \\
		& -              & -  & -    & qv\_n16\_d5  & 29 & 30.96 \\
		& \textbf{Sum}   & \textbf{34} & \textbf{2.84} & \textbf{Sum} & \textbf{56} & \textbf{40.73} \\
		\midrule
		\multirow{4}{*}{\rotatebox{90}{\textbf{Minor}}} 
		& ising\_model\_10 & 20  & 3.26  & ising\_model\_13 & 20  & 5.61  \\
		& qv\_n12\_d10     & 22  & 10.91 & ising\_model\_16 & 37  & 29.28 \\
		& rd84\_142        & 66  & 47.72 & qft\_16          & 55  & 61.16 \\
		& \textbf{Sum}     & \textbf{108} & \textbf{61.89} & \textbf{Sum} & \textbf{112} & \textbf{96.05} \\
		\midrule
		\multirow{6}{*}{\rotatebox{90}{\textbf{Moderate}}} 
		& sf\_276     & 301 & 19.04  & sf\_274     & 300 & 19.14  \\
		& rd53\_130   & 91  & 39.32  & con1\_216   & 212 & 81.75  \\
		& f2\_232     & 248 & 89.76  & wim\_266    & 138 & 171.85 \\
		& rd53\_251   & 206 & 129.11 & cm152a\_212 & 177 & 276.03 \\
		& -           & -   & -      & hwb5\_53    & 170 & 105.08 \\
		& \textbf{Sum} & \textbf{846} & \textbf{277.23} & \textbf{Sum} & \textbf{997} & \textbf{653.85} \\
		\midrule
		\multirow{3}{*}{\rotatebox{90}{\textbf{Major}}} 
		& pm1\_249    & 634  & 222.42 & cm42a\_207 & 634  & 224.51 \\
		& squar5\_261 & 459  & 408.11 & dc1\_220   & 443  & 296.86 \\
		& \textbf{Sum} & \textbf{1093} & \textbf{630.53} & \textbf{Sum} & \textbf{1077} & \textbf{521.37} \\
		\midrule
		\multirow{5}{*}{\rotatebox{90}{\textbf{Maximal}}} 
		& z4\_268         & 1113 & 303.20  & radd\_250       & 1212 & 426.40  \\
		& misex1\_241     & 1316 & 1549.20 & adr4\_197       & 721  & 872.70  \\
		& rd73\_252       & 935  & 1351.90 & sym6\_145       & 650  & 563.90  \\
		& cycle10\_2\_110 & 834  & 2347.10 & square\_root\_7 & 1475 & 3385.30 \\
		& \textbf{Sum}    & \textbf{4198} & \textbf{5551.50} & \textbf{Sum} & \textbf{4058} & \textbf{5248.50} \\
		\bottomrule
	\end{tabular}
	\begin{tablenotes}
		\footnotesize
		\item $^1$ Each QPU column contains the circuits assigned to that QPU, along with their compilation time $\Delta \mathrm{T}_{\mathrm{DYNAMO}}$ and the increase of post-compilation Rydberg stage counts $\Delta \mathrm{L}_{\mathrm{DYNAMO}}$.
	\end{tablenotes}
	
\end{table*}

Finally, we evaluate DYNAMO's ability to distribute quantum compilation tasks across multiple quantum resources. For the case of two quantum resources, the results for different circuit groups are shown in Table~\ref{tab:exp2-qpu-all}, where the resources are labeled as QPU 0 and QPU 1. $\Delta \text{L}_\text{DYNAMO}$ denotes the number of Rydberg stages required for executing the assigned circuits on each resource, and $\Delta \text{T}_\text{DYNAMO}$ is the total compilation time.

Across all groups from Minimal to Maximal, the $\Delta \text{L}_\text{DYNAMO}$ values on QPU 0 and QPU 1 remain closely aligned, indicating that DYNAMO effectively balances workloads across quantum resources. This load balancing helps maximize hardware utilization and minimize execution depth. Since compilation tasks on different resources are independent, we also apply multi-threading on the classical side to further reduce overall compilation time.

\begin{table*}[h]
	\footnotesize
	\caption{Multi-Program Compilation Results on Three Quantum Resources}
	\label{tab:exp2-qpu3-all}
	\centering
	\tabcolsep=0.115\linewidth
	\renewcommand{\arraystretch}{1.2}
	\begin{tabular}{@{}l@{\hspace{1.5em}}l@{\hspace{0.5em}}r@{\hspace{0.5em}}r@{\hspace{1em}}l@{\hspace{0.5em}}r@{\hspace{0.5em}}r@{\hspace{1em}}l@{\hspace{0.5em}}r@{\hspace{0.5em}}r@{}}
		\toprule
		& 
		\multicolumn{3}{c}{\textbf{QPU 0}} & 
		\multicolumn{3}{c}{\textbf{QPU 1}} & 
		\multicolumn{3}{c}{\textbf{QPU 2}} \\
		\cmidrule(lr){2-4}\cmidrule(lr){5-7}\cmidrule(lr){8-10}
		& \textbf{Circuit} & \textbf{$\Delta L$} & \textbf{$\Delta T$} &
		\textbf{Circuit} & \textbf{$\Delta L$} & \textbf{$\Delta T$} &
		\textbf{Circuit} & \textbf{$\Delta L$} & \textbf{$\Delta T$} \\
		\midrule
		\multirow{4}{*}{\rotatebox{90}{\textbf{Minimal}}} 
		& 4mod5-v1\_22 & 10 & 0.42  & bv\_n16   & 15 & 6.71  & alu-v0\_27     & 15 & 0.61 \\
		& qv\_n16\_d5  & 13 & 10.45 & 4gt13\_92 & 18 & 1.81  & mod5mils\_65   & 3  & 0.92 \\
		& -            & -  & -     & -         & -  & -     & decod24-v2\_43 & 12 & 1.16 \\
		& \textbf{Sum} & \textbf{23} & \textbf{10.87} & \textbf{Sum} & \textbf{33} & \textbf{8.52} & \textbf{Sum} & \textbf{30} & \textbf{2.69} \\
		\midrule
		\multirow{3}{*}{\rotatebox{90}{\textbf{Minor}}} 
		& ising\_model\_10 & 20 & 3.26  & ising\_model\_13 & 20 & 5.61  & ising\_model\_16 & 20 & 8.80  \\
		& qft\_16          & 51 & 34.27 & rd84\_142        & 75 & 39.33 & qv\_n12\_d10     & 38 & 16.34 \\
		& \textbf{Sum}     & \textbf{71} & \textbf{37.53} & \textbf{Sum} & \textbf{95} & \textbf{44.94} & \textbf{Sum} & \textbf{58} & \textbf{25.14} \\
		\midrule
		\multirow{4}{*}{\rotatebox{90}{\textbf{Moderate}}} 
		& sf\_276   & 301 & 19.04  & sf\_274     & 300 & 19.14  & con1\_216 & 346 & 47.89  \\
		& rd53\_251 & 200 & 65.06  & f2\_232     & 228 & 68.99  & wim\_266  & 188 & 136.94 \\
		& hwb5\_53  & 309 & 71.67  & cm152a\_212 & 230 & 238.56 & rd53\_130 & 149 & 82.96  \\
		& \textbf{Sum} & \textbf{810} & \textbf{155.77} & \textbf{Sum} & \textbf{758} & \textbf{326.69} & \textbf{Sum} & \textbf{683} & \textbf{267.79} \\
		\midrule
		\multirow{3}{*}{\rotatebox{90}{\textbf{Major}}} 
		& pm1\_249 & 634 & 222.42 & cm42a\_207 & 634 & 224.52 & dc1\_220    & 705  & 158.01 \\
		& -        & -   & -      & -          & -   & -      & squar5\_261 & 424  & 420.44 \\
		& \textbf{Sum} & \textbf{634} & \textbf{222.42} & \textbf{Sum} & \textbf{634} & \textbf{224.52} & \textbf{Sum} & \textbf{1129} & \textbf{578.45} \\
		\midrule
		\multirow{4}{*}{\rotatebox{90}{\textbf{Maximal}}} 
		& z4\_268         & 1113 & 303.27  & radd\_250       & 1212 & 426.46  & adr4\_197   & 1258 & 467.96  \\
		& square\_root\_7 & 2069 & 2402.18 & rd73\_252       & 1443 & 1003.50 & sym6\_145   & 944  & 433.62  \\
		& -               & -    & -       & cycle10\_2\_110 & 1139 & 1906.83 & misex1\_241 & 852  & 1903.77 \\
		& \textbf{Sum}    & \textbf{3182} & \textbf{2705.45} & \textbf{Sum} & \textbf{3794} & \textbf{3336.79} & \textbf{Sum} & \textbf{3054} & \textbf{2805.35} \\
		\bottomrule
	\end{tabular}
\end{table*}

We extend the experiment to three quantum resources (QPU 0, QPU 1, and QPU 2), with results presented in Table~\ref{tab:exp2-qpu3-all}. The results show that DYNAMO maintains strong workload balance and scheduling effectiveness even in more complex scenarios, confirming its robustness and adaptability to systems with increasing quantum capacity. The successful scaling from single-resource to multi-resource scenarios indicates that DYNAMO effectively addresses the global nature of AOD movement constraints that traditionally prevent simple resource partitioning approaches. These results support the broader vision of quantum operating systems by showing that DYNAMO can coordinate complex multi-resource quantum workloads while maintaining compilation efficiency and correctness.

The comprehensive experimental evaluation across three distinct scenarios collectively validates the fundamental thesis underlying this work: the transition from single-circuit compilation to multi-programming capabilities represents both a feasible and beneficial evolution for neutral atom quantum computing systems. The progression from pairwise to grouped to multi-resource compilation scenarios demonstrates that DYNAMO successfully bridges the identified gap between the architectural advantages of neutral atom systems and the practical requirements for quantum operating systems.

\section{Discussion}

Previous approaches to neutral atom quantum compilation have focused exclusively on single-circuit optimization, failing to exploit the inherent parallelism advantages of neutral atom architectures for multi-programming applications. DYNAMO addresses this limitation by enabling concurrent quantum program execution through spatial deformation modeling and constraint-based scheduling. The experimental results demonstrate substantial performance improvements, with compilation speedups of up to $14.39\times$ and execution stage reductions averaging $50.47\%$ across diverse circuit configurations. Most significantly, the performance advantages become increasingly pronounced with larger circuit groups and multi-resource scenarios, indicating that DYNAMO's effectiveness scales with system complexity.

The experimental evaluation confirms that DYNAMO successfully addresses the core challenges identified in our motivation: exploiting neutral atom parallelism while managing global AOD movement constraints that traditionally prevented resource partitioning strategies. The balanced workload distribution achieved across multiple quantum processing units validates the feasibility of implementing multi-programming capabilities on neutral atom platforms, establishing a critical foundation for quantum operating systems.

The current implementation employs heuristic optimization strategies that prioritize computational efficiency and scalability over global optimality, aligning with practical requirements of real-world quantum computing deployments where rapid compilation and resource responsiveness are essential. This design philosophy reflects common considerations in quantum system research, where foundational functionality establishment typically guides initial development phases.

Future research directions offer promising opportunities for extending DYNAMO's impact beyond neutral atom systems. The modular design suggests strong potential for adaptation to other quantum architectures, enabling multi-programming capabilities across diverse platforms. By establishing multi-programming as a foundational capability, this work contributes essential building blocks for comprehensive quantum operating systems.

\section{Conclusion}

Current quantum compilation methods support only single-circuit execution, creating a critical gap for quantum operating systems that require multi-programming capabilities. To address this challenge, we propose DYNAMO, a dynamic neutral atom multi-programming optimizer that enables concurrent quantum program execution. Our approach uses spatial deformation modeling and constraint-based scheduling to decompose multi-programming challenges into manageable problems through parallel compilation frameworks and cycle-wise decomposition techniques. Experimental evaluation demonstrates performance improvements with up to $14.39\times$ compilation speedup and $50.47\%$ average execution stage reduction while achieving balanced workload distribution across multiple quantum processing units.

Our investigation reveals important insights into neutral atom quantum systems for multi-programming applications. The inherent parallelism of neutral atom architectures provides natural opportunities for multi-programming that extend beyond traditional single-circuit approaches. Treating compiled circuits as dynamic space occupations enables effective resource sharing while managing complex AOD movement constraints, transforming global coordination challenges into structured spatial-temporal optimization problems. By enabling concurrent program execution, DYNAMO provides essential infrastructure for transitioning from single-application quantum devices to multi-tasking quantum computing systems, addresses a core challenge in quantum operating systems. Our work establishing neutral atom architectures as viable candidates for quantum operating system deployment. An important direction for future research lies in assessing the parallelization potential of next-generation quantum computing architectures, to further advance the scalability and practicality of multi-program compilation.

\section{Acknowledgements}
This work was supported in part by the National Natural Science Foundation of China under Grant 62472072, and in part by the National Natural Science Foundation of China under Grant 62172075.

\section*{Appendixes}

\subsection*{A.  Neutral Atom Hardware Compilation Process}

\label{Detailed Compilation}

This appendix provides detailed illustrations of neutral atom hardware mechanisms and compilation processes that support the foundational concepts presented in Section ~\ref{Preliminaries}. Through specific examples and visual demonstrations, we explain how the three fundamental compilation constraints affect circuit mapping and the strategies required to satisfy them.

In neutral atom quantum computing, individual atoms serve as qubits, held in place and manipulated using precisely controlled optical fields\cite{Baker2021}. The system employs two complementary types of traps: static traps created by a spatial light modulator (SLM) arrange qubits in a fixed lattice, while mobile traps generated by a two-dimensional acousto-optic deflector (AOD) enable dynamic movement of qubits within the plane. The AOD traps arise at the intersections of controllable rows and columns, allowing flexible reconfiguration by shifting these lines without altering their relative order — this constraint prevents collisions and loss of qubits during movement\cite{Tan2022,Tan2024a}. By aligning the mobile AOD traps with the static SLM traps and adjusting trap intensities, qubits can be transferred seamlessly between the two kind of traps. A key mechanism for entangling qubits involves a global Rydberg laser that excites atoms simultaneously, inducing multi-qubit interactions within a limited range defined by the blockade radius $r_b$. Two-qubit gates are realized only when pairs of qubits are brought within this interaction range, making the controlled relocation of qubits essential for executing quantum circuits. The computation proceeds in discrete steps known as Rydberg stages, during which parallel two-qubit gates operate on qubits occupying the same interaction sites. Between these stages, the flexible movement of qubits allows them to be repositioned to new sites, enabling interactions with different partners over time\cite{Tan2025}.

\begin{figure}[t]
	\centering
	\includegraphics[width=1\columnwidth, trim=10.5cm 8.5cm 10.5cm 8.5cm, clip]{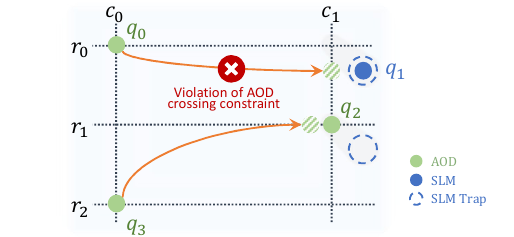}
	\caption{Illustration of AOD movement constraint.}
	\label{fig:na-compile0}
\end{figure}

\begin{figure}[t]
	\centering
	\includegraphics[width=1\columnwidth, trim=10.5cm 8.5cm 10.5cm 8.5cm, clip]{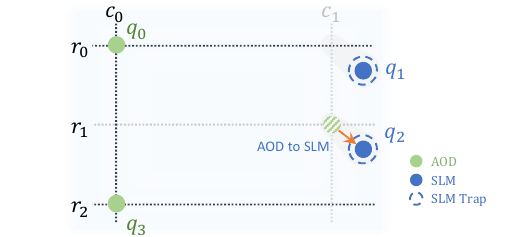}
	\caption{Qubit transfer.}
	\label{fig:na-compile1}
\end{figure}

\begin{figure}[t]
	\centering
	\includegraphics[width=1\columnwidth, trim=10.6cm 8.5cm 10.4cm 8.5cm, clip]{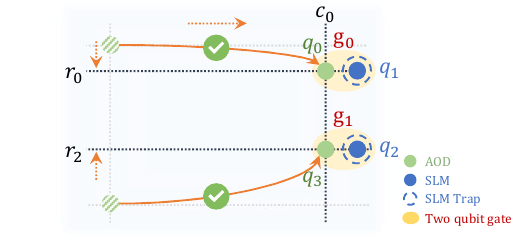}
	\caption{Realization of two parallel gates.}
	\label{fig:na-compile2}
\end{figure}

Bound by these guiding principles, the compilation process must carefully coordinate qubit movements while respecting hardware constraints. Figures \ref{fig:na-compile0} - \ref{fig:na-compile2} demonstrate how the above three constraints affect the compilation process.

In Figure \ref{fig:na-compile0}, there are two gates to be executed: $g_0$ operates on qubits $q_0$ and $q_1$, while $g_1$ operates on qubits $q_2$ and $q_3$. However, since $q_0$ and $q_1$, as well as $q_2$ and $q_3$, are positioned too far apart, they fail to satisfy the two-qubit gate execution constraint, rendering the gates unexecutable. To execute $g_0$ and $g_1$, $q_0$ and $q_3$ need to be moved to positions close to $q_1$ and $q_2$, respectively. However, moving $q_0$ close to $q_1$ would cause AOD column $c_0$ to cross over $c_1$, violating the AOD movement constraint and making this approach infeasible.

To address this problem, one solution is shown in Figure \ref{fig:na-compile1}, where $q_2$ in the AOD trap is transferred to an adjacent SLM trap. Since qubits in SLM traps do not require AOD tweezers for maintenance, $c_1$ no longer obstructs $c_0$.

Subsequently, as illustrated in Figure \ref{fig:na-compile2}, by moving $r_2$ and $c_0$, $q_0$ and $q_3$ are moved to positions adjacent to $q_1$ and $q_2$. Meanwhile, the distance between the two qubit pairs that need to execute two-qubit gates, $(q_0, q_1)$ (executing $g_0$) and $(q_2, q_3)$ (executing $g_1$), is sufficiently large, allowing gates $g_0$ and $g_1$ to be executed in parallel. 

Examples above illustrate the intricate coordination required in neutral atom compilation, where spatial positioning, temporal sequencing, and constraint satisfaction must be simultaneously optimized. The global nature of AOD movement effects and the complex interplay between hardware limitations create compilation challenges that become even more complex in multi-programming scenarios where multiple circuits must share the same hardware resources.

\subsection*{B.  Multi-programming Compilation Method Implementation Details}

\label{Multi-programming Compilation Details}

This appendix provides the technical implementation details, mathematical formulations, and variable definitions that support the multi-programming compilation method presented in \ref{Multi-programming Compilation Method}.

The mathematical formulation of method builds upon the variable definitions established in the DPQA framework \cite{Tan2024a}, as summarized in Table\ref{tab:variables}.

\begin{table}[t]
	\renewcommand{\arraystretch}{1}
	\caption{Variable definitions adopted from DPQA\cite{Tan2024a}.}
	\label{tab:variables}
	\centering
	\tabcolsep=0.02\linewidth
	\begin{tabular}{cc}
		\toprule
		Variable & Description \\
		\midrule
		$x_{i,t}$ & $x$-coordinate of qubit $i$ on the SLM plane at stage $t$ \\
		$y_{i,t}$ & $y$-coordinate of qubit $i$ on the SLM plane at stage $t$ \\
		$a_{i,t}$ & Binary indicator of trap type: $1$ for AOD, $0$ for SLM \\
		$c_{i,t}$ & Index of the AOD column used by qubit $i$ at stage $t$ \\
		$r_{i,t}$ & Index of the AOD row used by qubit $i$ at stage $t$ \\
		$t_j$ & Stage at which quantum gate $j$ is executed \\
		\bottomrule
	\end{tabular}
\end{table}

\begin{figure}[h]
	\centering
	\includegraphics[width=0.85\columnwidth, trim=11.5cm 8cm 11.5cm 8cm, clip]{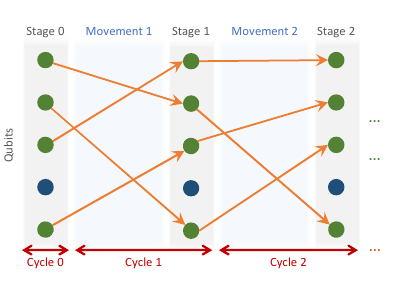}
	\caption{Comparasion of Cycle and Stage, Cycle contains extra movement information for compilation space deformation.}
	\label{fig:cycle-stage}
\end{figure}

In DPQA, the concept of a stage is introduced as a temporal abstraction to discretize the compilation timeline, where each stage captures the physical locations of all qubits. In contrast, we define a cycle as consisting of two steps: the AOD movement step and the non-movement operation step, which together describe both the intermediate relocation of qubits and the operations executed at each stage, as illustrated in Figure~\ref{fig:cycle-stage}.

Let the set of $n$ already compiled circuits on the neutral atom array be denoted by $C_s = \{ C_{s,1}, C_{s,2}, \dots, C_{s,n} \}$. For each compiled circuit, we apply the stage-wise decomposition discussed earlier to extract its execution sequence, yielding $O_s = \{ O_{s,1}, O_{s,2}, \dots, O_{s,n} \}$. Each $O_{s,j}$ contains the a serial of stage and step AOD movements and logical operations of the corresponding compiled circuit $C_{s,j}$. To compile a new circuit without violating the spatial constraints imposed by existing circuits, we define the following constraints:

\textbf{\textit{(a). AOD Movement Constraints}}

To prevent spatial conflicts during AOD movement steps, we impose constraints based on previously scheduled AOD operations in each stage. For any movement $M_{j,k,p}$ in compiled circuit execution sequence $O_{s,j}$ at cycle $k$, where $(x_{0,j,k,p}^s,y_{0,j,k,p}^s)$ and $(x_{1,j,k,p}^s,y_{1,j,k,p}^s)$ are start and end SLM coordinates of $M_{j,k,p}$, the following constraints must hold:

For horizontal AOD movement constraints,
\begin{align}
	\forall i,\ \forall j,\ \forall p,\ \forall t > 0,\quad x_{i,t-1} < x_{0,j,t,p}^s \ \Rightarrow\ x_{i,t} < x_{1,i,t,k}^s \label{eq:x-left}, \\
	\forall i,\ \forall j,\ \forall p,\ \forall t > 0,\quad x_{i,t-1} > x_{0,j,t,p}^s \ \Rightarrow\ x_{i,t} > x_{1,i,t,k}^s \label{eq:x-right}, \\
	\forall i,\ \forall j,\ \forall p,\ \forall t > 0,\quad x_{i,t-1} = x_{0,j,t,p}^s \ \Rightarrow\ x_{i,t} = x_{1,i,t,k}^s \label{eq:x-equ}.
\end{align}

For vertical AOD movement constraints,
\begin{align}
	\forall i,\ \forall k,\ \forall t > 0,\quad y_{i,t-1} < y_{0,i,t,k}^s \ \Rightarrow\ y_{i,t} < y_{1,i,t,k}^s \label{eq:y-left}, \\
	\forall i,\ \forall k,\ \forall t > 0,\quad y_{i,t-1} > y_{0,i,t,k}^s \ \Rightarrow\ y_{i,t} > y_{1,i,t,k}^s \label{eq:y-right}, \\
	\forall i,\ \forall k,\ \forall t > 0,\quad y_{i,t-1} = y_{0,i,t,k}^s \ \Rightarrow\ y_{i,t} = y_{1,i,t,k}^s. \label{eq:y-equ}
\end{align}

These rules guarantee that any new movement will not violate the directionality of existing AOD movements, thus maintaining the integrity of the Order-Preserving Zones defined by prior programs.

\textbf{\textit{(b). Two-Qubit Gate Parallelization Constraint}}

To ensure spatial exclusivity of two-qubit gates, we impose a constraint that prohibits any gate in the circuit being compiled from overlapping, at any stage, with gates already scheduled in previously compiled circuits. Formally, let $G_{j,k,l}^s$ denote the $l$-th two-qubit gate of $k$-th cycle in circuit $C_{s,j}$, and $(x_{j,k,l}^s, y_{j,k,l}^s)$ denotes the SLM coordinate of gate $G_{j,k,l}$, $(q_{0}^s, q_{1}^s)$ denotes the two qubits of gate $G_{j,k,l}$, the following constraint must hold:

\begin{equation}
	\forall j,\ \forall k,\ \forall l,\ \forall g,\quad 
	t_g = l \ \Rightarrow\ 
	\left( x_{q_{0}^s,l} \neq x_{j,k,l}^s \ \lor\ y_{q_{0}^s,l} \neq y_{j,k,l}^s \right),
	\label{eq:gate-paral-q0}
\end{equation}

\begin{equation}
	\forall j,\ \forall k,\ \forall l,\ \forall g,\quad 
	t_g = l \ \Rightarrow\ 
	\left( x_{q_{1}^s,l} \neq x_{j,k,l}^s \ \lor\ y_{q_{1}^s,l} \neq y_{j,k,l}^s \right).
	\label{eq:gate-paral-q}
\end{equation}

These constraints complete the mathematical formulation of the scheduling problem under multi-program interference. On top of this formulation, we design a constraint-based scheduler that incorporates array zoning and step placement logic to support concurrent compilation. Code available on \href{https://github.com/StillwaterQ/DYNAMO}{Github}.

\section{Experimental Implementation Details}
\label{Experimental Details}

Since DPQA natively supports only single-program compilation, we design two adaptation schemes to enable it to compile multiple quantum circuits for comparison purposes. These schemes provide comprehensive baseline coverage for evaluating DYNAMO's multi-programming capabilities.

\textbf{(a). Sequential Compilation Scheme ($\text{DPQA}^s$):} 

The sequential compilation scheme compiles multiple circuits one after another in a serial manner. We record the total compilation time and the cumulative stage count required to execute all circuits. This approach represents the straightforward extension of single-circuit compilation to multiple programs, providing a baseline for understanding the benefits of concurrent execution.

\textbf{(b). Merged Compilation Scheme ($\text{DPQA}^c$):} 

The merged compilation scheme combines all target circuits into a single quantum circuit before compilation. To prevent interference between different programs, the qubits from each circuit are kept logically independent. For example, if the input circuits are $qc_1$, $qc_2$ and $qc_3$, containing 4, 6, and 5 qubits respectively, the merged circuit will operate on 15 qubits: qubits 0–3 execute all operations from $qc_1$, qubits 4–9 execute those from $qc_2$, and qubits 10–14 execute those from $qc_3$. We record the total compilation time and resulting stage count. This scheme tests whether treating multiple programs as a single large circuit can achieve compilation efficiency, representing an alternative approach to multi-programming.

By evaluating both schemes, we establish a comprehensive baseline for assessing the compilation efficiency and effectiveness of our proposed DYNAMO framework in multi-programming scenarios.

All experiments were conducted using the same hardware and software configurations to ensure fair and reproducible comparisons. The hardware setup consists of an AMD Ryzen 9 7900X CPU running at 4.7GHz with 32GB of RAM. The software environment includes a Linux operating system, Python 3.9, Qiskit 1.0.0 for quantum circuit representation, and the Z3 SMT solver (version 4.12.5.0) for constraint solving. A time limit of 10,000 seconds was applied to each compilation task to ensure reasonable experimental duration while allowing sufficient time for complex compilation scenarios.

\newpage

\bibliographystyle{IEEEtran}
\bibliography{ref}

% Generated by IEEEtran.bst, version: 1.14 (2015/08/26)
\begin{thebibliography}{10}
\providecommand{\url}[1]{#1}
\csname url@samestyle\endcsname
\providecommand{\newblock}{\relax}
\providecommand{\bibinfo}[2]{#2}
\providecommand{\BIBentrySTDinterwordspacing}{\spaceskip=0pt\relax}
\providecommand{\BIBentryALTinterwordstretchfactor}{4}
\providecommand{\BIBentryALTinterwordspacing}{\spaceskip=\fontdimen2\font plus
\BIBentryALTinterwordstretchfactor\fontdimen3\font minus
  \fontdimen4\font\relax}
\providecommand{\BIBforeignlanguage}[2]{{%
\expandafter\ifx\csname l@#1\endcsname\relax
\typeout{** WARNING: IEEEtran.bst: No hyphenation pattern has been}%
\typeout{** loaded for the language `#1'. Using the pattern for}%
\typeout{** the default language instead.}%
\else
\language=\csname l@#1\endcsname
\fi
#2}}
\providecommand{\BIBdecl}{\relax}
\BIBdecl

\bibitem{Bernstein2017}
D.~J. Bernstein and T.~Lange, ``Post-quantum cryptography,'' \emph{Nature},
  vol. 549, no. 7671, pp. 188--194, Sep. 2017.

\bibitem{Lancellotti2024}
G.~Lancellotti, S.~Perriello, A.~Barenghi, and G.~Pelosi, ``Design of a
  {{Quantum Walk Circuit}} to {{Solve}} the {{Subset-Sum Problem}},'' in
  \emph{Proceedings of the 61st {{ACM}}/{{IEEE Design Automation Conference}}},
  ser. {{DAC}} '24.\hskip 1em plus 0.5em minus 0.4em\relax {New York, NY, USA}:
  {Association for Computing Machinery}, Nov. 2024, pp. 1--6.

\bibitem{Abbas2024}
A.~Abbas \emph{et~al.}, ``Challenges and opportunities in quantum
  optimization,'' \emph{Nature Reviews Physics}, vol.~6, no.~12, pp. 718--735,
  Dec. 2024.

\bibitem{Clinton2024}
L.~Clinton \emph{et~al.}, ``Towards near-term quantum simulation of
  materials,'' \emph{Nature Communications}, vol.~15, no.~1, p. 211, Jan. 2024.

\bibitem{Harrigan2021}
\BIBentryALTinterwordspacing
M.~P. Harrigan \emph{et~al.}, ``\BIBforeignlanguage{en}{Quantum approximate
  optimization of non-planar graph problems on a planar superconducting
  processor},'' \emph{\BIBforeignlanguage{en}{Nature Physics}}, vol.~17, no.~3,
  pp. 332--336, Mar. 2021. [Online]. Available:
  \url{https://www.nature.com/articles/s41567-020-01105-y}
\BIBentrySTDinterwordspacing

\bibitem{Arute2019}
\BIBentryALTinterwordspacing
F.~Arute \emph{et~al.}, ``\BIBforeignlanguage{en}{Quantum supremacy using a
  programmable superconducting processor},''
  \emph{\BIBforeignlanguage{en}{Nature}}, vol. 574, no. 7779, pp. 505--510,
  Oct. 2019, number: 7779 Publisher: Nature Publishing Group. [Online].
  Available: \url{https://www.nature.com/articles/s41586-019-1666-5}
\BIBentrySTDinterwordspacing

\bibitem{Huang2023}
X.~Huang, J.~Luo, and L.~Li, ``Quantum speedup and limitations on matroid
  property problems,'' \emph{Frontiers of Computer Science}, vol.~18, no.~4, p.
  184905, Dec. 2023.

\bibitem{Wu2013}
N.~Wu, H.~Hu, F.~Song, H.~Zheng, and X.~Li, ``Quantum software framework: A
  tentative study,'' \emph{Frontiers of Computer Science}, vol.~7, no.~3, pp.
  341--349, Jun. 2013.

\bibitem{corrigan2017quantum}
H.~Corrigan-Gibbs, D.~J. Wu, and D.~Boneh, ``Quantum operating systems,'' in
  \emph{Proceedings of the 16th Workshop on Hot Topics in Operating Systems},
  2017, pp. 76--81.

\bibitem{acharya2024quantum}
R.~Acharya \emph{et~al.}, ``Quantum error correction below the surface code
  threshold,'' \emph{Nature}, 2024.

\bibitem{Das2019}
P.~Das, S.~S. Tannu, P.~J. Nair, and M.~Qureshi, ``A {{Case}} for
  {{Multi-Programming Quantum Computers}},'' in \emph{Proceedings of the 52nd
  {{Annual IEEE}}/{{ACM International Symposium}} on {{Microarchitecture}}},
  ser. {{MICRO}} '52.\hskip 1em plus 0.5em minus 0.4em\relax {New York, NY,
  USA}: {Association for Computing Machinery}, Oct. 2019, pp. 291--303.

\bibitem{niu2023enabling}
S.~Niu and A.~Todri-Sanial, ``Enabling multi-programming mechanism for quantum
  computing in the nisq era,'' \emph{Quantum}, vol.~7, p. 925, 2023.

\bibitem{Liu2021}
L.~Liu and X.~Dou, ``{{QuCloud}}: {{A New Qubit Mapping Mechanism}} for
  {{Multi-programming Quantum Computing}} in {{Cloud Environment}},'' in
  \emph{2021 {{IEEE International Symposium}} on {{High-Performance Computer
  Architecture}} ({{HPCA}})}, Feb. 2021, pp. 167--178.

\bibitem{Ohkura2022}
Y.~Ohkura, T.~Satoh, and R.~Van~Meter, ``Simultaneous {{Execution}} of
  {{Quantum Circuits}} on {{Current}} and {{Near-Future NISQ Systems}},''
  \emph{IEEE Transactions on Quantum Engineering}, vol.~3, pp. 1--10, 2022.

\bibitem{Liu2024}
L.~Liu and X.~Dou, ``{{QuCloud}}+: {{A Holistic Qubit Mapping Scheme}} for
  {{Single}}/{{Multi-programming}} on {{2D}}/{{3D NISQ Quantum Computers}},''
  \emph{ACM Trans. Archit. Code Optim.}, vol.~21, no.~1, pp. 9:1--9:27, Jan.
  2024.

\bibitem{Orenstein2024}
A.~Orenstein and V.~Chaudhary, ``{{QGroup}}: {{Parallel Quantum Job Scheduling
  Using Dynamic Programming}},'' in \emph{2024 {{IEEE International
  Conference}} on {{Quantum Computing}} and {{Engineering}} ({{QCE}})},
  vol.~01, Sep. 2024, pp. 990--999.

\bibitem{Li2025}
J.~Li, Y.~Song, Y.~Liu, J.~Pan, L.~Yang, T.~Humble, and W.~Jiang,
  ``{{QuSplit}}: {{Achieving Both High Fidelity}} and {{Throughput}} via {{Job
  Splitting}} on {{Noisy Quantum Computers}},'' Mar. 2025.

\bibitem{Evered2023}
S.~J. Evered \emph{et~al.}, ``High-fidelity parallel entangling gates on a
  neutral-atom quantum computer,'' \emph{Nature}, vol. 622, no. 7982, pp.
  268--272, Oct. 2023.

\bibitem{Sunami2025}
S.~Sunami, S.~Tamiya, R.~Inoue, H.~Yamasaki, and A.~Goban, ``Scalable
  {{Networking}} of {{Neutral-Atom Qubits}}: {{Nanofiber-Based Approach}} for
  {{Multiprocessor Fault-Tolerant Quantum Computers}},'' \emph{PRX Quantum},
  vol.~6, no.~1, p. 010101, Feb. 2025.

\bibitem{Henriet2020}
L.~Henriet, L.~Beguin, A.~Signoles, T.~Lahaye, A.~Browaeys, G.-O. Reymond, and
  C.~Jurczak, ``Quantum computing with neutral atoms,'' \emph{Quantum}, vol.~4,
  p. 327, Sep. 2020.

\bibitem{Ebadi2021}
S.~Ebadi \emph{et~al.}, ``Quantum phases of matter on a 256-atom programmable
  quantum simulator,'' \emph{Nature}, vol. 595, no. 7866, pp. 227--232, Jul.
  2021.

\bibitem{Bluvstein2022}
D.~Bluvstein \emph{et~al.}, ``A quantum processor based on coherent transport
  of entangled atom arrays,'' \emph{Nature}, vol. 604, no. 7906, pp. 451--456,
  Apr. 2022.

\bibitem{Bluvstein2024}
------, ``Logical quantum processor based on reconfigurable atom arrays,''
  \emph{Nature}, vol. 626, no. 7997, pp. 58--65, Feb. 2024.

\bibitem{Baker2021}
\BIBentryALTinterwordspacing
J.~M. Baker \emph{et~al.}, ``Exploiting long-distance interactions and
  tolerating atom loss in neutral atom quantum architectures,'' in
  \emph{Proceedings of the 48th {Annual} {International} {Symposium} on
  {Computer} {Architecture}}, ser. {ISCA} '21.\hskip 1em plus 0.5em minus
  0.4em\relax Virtual Event, Spain: IEEE Press, Nov. 2021, pp. 818--831.
  [Online]. Available:
  \url{https://dl.acm.org/doi/10.1109/ISCA52012.2021.00069}
\BIBentrySTDinterwordspacing

\bibitem{Li2023}
\BIBentryALTinterwordspacing
Y.~Li, Y.~Zhang, M.~Chen, X.~Li, and P.~Xu, ``Timing-{Aware} {Qubit} {Mapping}
  and {Gate} {Scheduling} {Adapted} to {Neutral} {Atom} {Quantum}
  {Computing},'' \emph{IEEE Transactions on Computer-Aided Design of Integrated
  Circuits and Systems}, vol.~42, no.~11, pp. 3768--3780, Nov. 2023, conference
  Name: IEEE Transactions on Computer-Aided Design of Integrated Circuits and
  Systems. [Online]. Available:
  \url{https://ieeexplore.ieee.org/document/10082942}
\BIBentrySTDinterwordspacing

\bibitem{Patel2023}
T.~Patel, D.~Silver, and D.~Tiwari, ``{{GRAPHINE}}: {{Enhanced Neutral Atom
  Quantum Computing}} using {{Application-Specific Rydberg Atom
  Arrangement}},'' in \emph{Proceedings of the {{International Conference}} for
  {{High Performance Computing}}, {{Networking}}, {{Storage}} and
  {{Analysis}}}, ser. {{SC}} '23.\hskip 1em plus 0.5em minus 0.4em\relax {New
  York, NY, USA}: {Association for Computing Machinery}, Nov. 2023, pp. 1--15.

\bibitem{Tan2022}
B.~Tan, D.~Bluvstein, M.~D. Lukin, and J.~Cong, ``Qubit {{Mapping}} for
  {{Reconfigurable Atom Arrays}},'' in \emph{Proceedings of the 41st
  {{IEEE}}/{{ACM International Conference}} on {{Computer-Aided
  Design}}}.\hskip 1em plus 0.5em minus 0.4em\relax {San Diego California}:
  {ACM}, Oct. 2022, pp. 1--9.

\bibitem{Tan2024}
D.~B. Tan, W.-H. Lin, and J.~Cong, ``Compilation for {{Dynamically
  Field-Programmable Qubit Arrays}} with {{Efficient}} and {{Provably
  Near-Optimal Scheduling}},'' Nov. 2024.

\bibitem{Ludmir2024}
J.~Ludmir and T.~Patel, ``Parallax: {{A Compiler}} for {{Neutral Atom Quantum
  Computers}} under {{Hardware Constraints}},'' Oct. 2024.

\bibitem{Khatri2019}
S.~Khatri, R.~LaRose, A.~Poremba, L.~Cincio, A.~T. Sornborger, and P.~J. Coles,
  ``Quantum-assisted quantum compiling,'' \emph{Quantum}, vol.~3, p. 140, May
  2019.

\bibitem{Guo2024}
Z.-H. Guo and T.-C. Wang, ``{{SMT-Based Layout Synthesis Approaches}} for
  {{Quantum Circuits}},'' in \emph{Proceedings of the 2024 {{International
  Symposium}} on {{Physical Design}}}, ser. {{ISPD}} '24.\hskip 1em plus 0.5em
  minus 0.4em\relax {New York, NY, USA}: {Association for Computing Machinery},
  Mar. 2024, pp. 235--243.

\bibitem{Lin2023}
\BIBentryALTinterwordspacing
W.-H. Lin, J.~Kimko, B.~Tan, N.~Bjørner, and J.~Cong, ``Scalable {Optimal}
  {Layout} {Synthesis} for {NISQ} {Quantum} {Processors},'' in \emph{2023 60th
  ACM/IEEE Des. Autom. Conf. (DAC)}, Jul. 2023, pp. 1--6. [Online]. Available:
  \url{https://ieeexplore.ieee.org/document/10247760}
\BIBentrySTDinterwordspacing

\bibitem{Tan2024a}
B.~Tan, D.~Bluvstein, M.~Lukin, and J.~Cong, ``Compiling {{Quantum Circuits}}
  for {{Dynamically Field-Programmable Neutral Atoms Array Processors}},''
  \emph{Quantum}, vol.~8, p. 1281, Mar. 2024.

\bibitem{Lin2024}
W.-H. Lin, D.~B. Tan, and J.~Cong, ``Reuse-{{Aware Compilation}} for {{Zoned
  Quantum Architectures Based}} on {{Neutral Atoms}},'' Dec. 2024.

\bibitem{Tan2025}
D.~B. Tan, W.-H. Lin, and J.~Cong, ``Compilation for dynamically
  field-programmable qubit arrays with efficient and provably near-optimal
  scheduling,'' in \emph{Proceedings of the 30th Asia and South Pacific Design
  Automation Conference}.\hskip 1em plus 0.5em minus 0.4em\relax {New York, NY,
  USA}: {Association for Computing Machinery}, 2025, pp. 921--929.

\bibitem{Tan2020a}
B.~Tan and J.~Cong, ``Optimal {{Layout Synthesis}} for {{Quantum Computing}},''
  in \emph{2020 {{IEEE}}/{{ACM International Conference On Computer Aided
  Design}} ({{ICCAD}})}, Nov. 2020, pp. 1--9.

\bibitem{Tan2021}
\BIBentryALTinterwordspacing
------, ``Optimal qubit mapping with simultaneous gate absorption,'' in
  \emph{2021 {IEEE}/{ACM} Int. Conf. Comput.-Aided Des. (ICCAD)}, Nov. 2021,
  pp. 1--8. [Online]. Available:
  \url{https://ieeexplore.ieee.org/document/9643554}
\BIBentrySTDinterwordspacing

\bibitem{Molavi2022}
\BIBentryALTinterwordspacing
A.~Molavi, A.~Xu, M.~Diges, L.~Pick, S.~Tannu, and A.~Albarghouthi, ``Qubit
  {Mapping} and {Routing} via {MaxSAT},'' in \emph{2022 55th IEEE/ACM Int.
  Symp. on {Microarchitecture} ({MICRO})}, Oct. 2022, pp. 1078--1091. [Online].
  Available: \url{https://ieeexplore.ieee.org/document/9923822}
\BIBentrySTDinterwordspacing

\bibitem{Cong2023}
\BIBentryALTinterwordspacing
J.~Cong, ``Lightning {Talk}: {Scaling} {Up} {Quantum} {Compilation} –
  {Challenges} and {Opportunities},'' in \emph{2023 60th ACM/IEEE Des. Autom.
  Conf. (DAC)}, Jul. 2023, pp. 1--2. [Online]. Available:
  \url{https://ieeexplore.ieee.org/document/10247677}
\BIBentrySTDinterwordspacing

\bibitem{Shaik2024}
\BIBentryALTinterwordspacing
I.~Shaik and J.~van~de Pol, ``\BIBforeignlanguage{en}{Optimal {Layout}
  {Synthesis} for {Deep} {Quantum} {Circuits} on {NISQ} {Processors} with 100+
  {Qubits}},'' in
  \emph{\BIBforeignlanguage{en}{{DROPS}-{IDN}/v2/document/10.4230/{LIPIcs}.{SAT}.2024.26}},
  2024. [Online]. Available:
  \url{https://drops.dagstuhl.de/entities/document/10.4230/LIPIcs.SAT.2024.26}
\BIBentrySTDinterwordspacing

\bibitem{Pattnaik2016}
A.~Pattnaik \emph{et~al.}, ``Scheduling {{Techniques}} for {{GPU
  Architectures}} with {{Processing-In-Memory Capabilities}},'' in
  \emph{Proceedings of the 2016 {{International Conference}} on {{Parallel
  Architectures}} and {{Compilation}}}, ser. {{PACT}} '16.\hskip 1em plus 0.5em
  minus 0.4em\relax {New York, NY, USA}: {Association for Computing Machinery},
  Sep. 2016, pp. 31--44.

\bibitem{Margiolas2016}
C.~Margiolas and M.~F.~P. O'Boyle, ``Portable and transparent software managed
  scheduling on accelerators for fair resource sharing,'' in \emph{Proceedings
  of the 2016 {{International Symposium}} on {{Code Generation}} and
  {{Optimization}}}, ser. {{CGO}} '16.\hskip 1em plus 0.5em minus 0.4em\relax
  {New York, NY, USA}: {Association for Computing Machinery}, Feb. 2016, pp.
  82--93.

\bibitem{Tillenius2015}
M.~Tillenius, E.~Larsson, R.~M. Badia, and X.~Martorell, ``Resource-{{Aware
  Task Scheduling}},'' \emph{ACM Trans. Embed. Comput. Syst.}, vol.~14, no.~1,
  pp. 5:1--5:25, Jan. 2015.

\bibitem{Rey2016}
A.~Rey, F.~D. Igual, and M.~{Prieto-Mat{\'i}as}, ``{{HeSP}}: {{A Simulation
  Framework}} for {{Solving}} the~{{Task Scheduling-Partitioning Problem}}
  on~{{Heterogeneous Architectures}},'' in \emph{Euro-{{Par}} 2016: {{Parallel
  Processing}}}, P.-F. Dutot and D.~Trystram, Eds.\hskip 1em plus 0.5em minus
  0.4em\relax {Cham}: {Springer International Publishing}, 2016, pp. 183--195.

\bibitem{CohenMaxime2017}
C.~C, KellerPhilipp, MirrokniVahab, and ZadimoghadddamMorteza, ``Overcommitment
  in {{Cloud Services Bin}} packing with {{Chance Constraints}},'' \emph{ACM
  SIGMETRICS Performance Evaluation Review}, Jun. 2017.

\bibitem{Sivarajah2020}
S.~Sivarajah, S.~Dilkes, A.~Cowtan, W.~Simmons, A.~Edgington, and R.~Duncan,
  ``T|ket{$\rangle$}: A retargetable compiler for {{NISQ}} devices,''
  \emph{Quantum Science and Technology}, vol.~6, no.~1, p. 014003, Nov. 2020.

\bibitem{Li2019}
G.~Li, Y.~Ding, and Y.~Xie, ``Tackling the {{Qubit Mapping Problem}} for
  {{NISQ-Era Quantum Devices}},'' in \emph{Proceedings of the {{Twenty-Fourth
  International Conference}} on {{Architectural Support}} for {{Programming
  Languages}} and {{Operating Systems}}}, ser. {{ASPLOS}} '19.\hskip 1em plus
  0.5em minus 0.4em\relax {New York, NY, USA}: {Association for Computing
  Machinery}, Apr. 2019, pp. 1001--1014.

\bibitem{Wille2008}
R.~Wille, D.~Gro{\ss}e, L.~Teuber, G.~W. Dueck, and R.~Drechsler, ``{{RevLib}}:
  {{An Online Resource}} for {{Reversible Functions}} and {{Reversible
  Circuits}},'' in \emph{38th {{International Symposium}} on {{Multiple Valued
  Logic}} (Ismvl 2008)}, May 2008, pp. 220--225.

\end{thebibliography}

\newpage

\vspace{1.5em}  % ← 增加垂直空间（可调节）

\end{document}